\documentclass[preprintnumbers,10pt,nofootinbib]{revtex4}

\usepackage{amsmath,latexsym,amssymb,amsfonts}
\usepackage[dvips]{graphicx,color}
\usepackage{bm}

\begin{document}


\title{\textbf{Page curves for tripartite systems}}

\author{
{\textsc{Junha Hwang$^{a}$}}, 
{\textsc{Deok Sang Lee$^{a}$}}, 
{\textsc{Dongju Nho$^{a}$}},\\ 
{\textsc{Jeonghun Oh$^{a}$}}, 
{\textsc{Hyosub Park$^{a}$}}, 
{\textsc{Dong-han Yeom$^{b}$}\footnote{{\tt innocent.yeom{}@{}gmail.com}}}
and 
{\textsc{Heeseung Zoe$^{a}$}\footnote{{\tt heezoe{}@{}dgist.ac.kr}}}
}

\affiliation{
$^{a}$\small{School of Undergraduate Studies, College of Transdisciplinary Studies,\\  
Daegu Gyeongbuk Institute of Science and Technology (DGIST), Daegu 42988, Republic of Korea}\\
$^{b}$\small{Leung Center for Cosmology and Particle Astrophysics, National Taiwan University, Taipei 10617, Taiwan}
}

\begin{abstract}
We investigate information flow and Page curves for tripartite systems. We prepare a tripartite system (say, $A$, $B$, and $C$) of a given number of states and calculate information and entropy contents by assuming random states. Initially, every particle was in $A$ (this means a black hole), and as time goes on, particles move to either $B$ (means Hawking radiation) or $C$ (means a broadly defined remnant, including a non-local transport of information, the last burst, an interior large volume, or a bubble universe, etc.). If the final number of states of the remnant is smaller than that of Hawking radiation, then information will be stored by both of the radiation and the mutual information between the radiation and the remnant, while the remnant itself does not contain information. On the other hand, if the final number of states of the remnant is greater than that of Hawking radiation, then the radiation contains negligible information, while the remnant and the mutual information between the radiation and the remnant contain information. Unless the number of states of the remnant is large enough compared to the entropy of the black hole, Hawking radiation \textit{must} contain information; and we meet the menace of black hole complementarity again. Therefore, this contrasts the tension between various assumptions and candidates of the resolution of the information loss problem.
\end{abstract}

\maketitle

\newpage

\tableofcontents

\section{Introduction}

The information loss problem \cite{Hawking:1976ra} is the interesting unresolved problem that is related to general relativity, quantum field theory, and the quantum theory of gravity. If there is a consistent theory of quantum gravity, then the information should be conserved for any processes including a black hole evaporation \cite{Hawking:1974sw}. On the other hand, if we cannot explain that the black hole evaporation is no more unitary, then we need to rethink about the meaning of the quantum theory of gravity.

We do not have a consensus on the quantum theory of gravity that may give the solution to this problem, but we can only list various possibilities and candidates of the resolution \cite{Ong:2016iwi}.

The simplest way to resolve the problem is to think that information is attached by Hawking radiation. However, if this is the case, it may imply that information should be duplicated between inside and outside the black hole as black hole complementarity argued \cite{Susskind:1993if}. However, it is also known that the original assumptions of black hole complementarity cannot be true \cite{Yeom:2008qw}, and hence we need to drop one of the assumptions of black hole complementarity, e.g., dropping general relativity and introducing the firewall \cite{Almheiri:2012rt}. On the other hand, if we introduce the firewall at once, then it should also modify not only the interior of the black hole but also the asymptotic future infinity \cite{Hwang:2012nn}.

Following these inconsistency arguments of black hole complementarity, if we assume that Hawking radiation does not contain information, then perhaps, a more conservative resolution is to think that the information is retained by an object, e.g., a remnant. There are various versions of this resolution, where we can briefly summarize this as the remnant picture \cite{Chen:2014jwq}. This picture can include various assertions, e.g., the last burst of the black hole evaporation contains all information \cite{Ashtekar:2005cj}, the Planck scale object carries all information \cite{Adler:2001vs}, entanglements between inside and outside regions contains information \cite{Horowitz:2003he,Smolin:2005tz}, an internal large volume carries information \cite{Christodoulou:2014yia}, or a bubble universe inside the black hole contains all information \cite{Farhi:1989yr}.

What will be the correct answer? In order to decide it, we need to check the consistency of each idea. In other words, what will be the condition that each idea (either Hawking radiation carries information or a remnant carries information) works to resolve the information loss problem?

In order to check the consistency of each idea, one interesting way to quantitative analysis was developed by Page \cite{Page:1993df,Page:1993wv}. We introduce a closed system with a fixed number of states and divide it to two subsystems; and bring particles from one part (black hole) to the other part (Hawking radiation). Then we can trace the flow of information; we call this information curve as the \textit{Page curve}. In these works, the people could use an averaged formula of the entropy of subsystems for a pure and random state \cite{Page:1993df}. 

In this paper, by using numerical techniques, we want to investigate the (broadly defined) remnant picture. In order to do this, we introduce one part more (the remnant part); hence, now we are dealing with a \textit{tripartite system} (a similar trial was done in \cite{Alonso-Serrano:2015bcr} in the context of the last burst picture \cite{Ashtekar:2005cj}). In this system, we can define various kinds of information, e.g., information of each part \cite{Lloyd:1988cn} as well as mutual information, and also trace their variations as the black hole evaporates. In order to calculate the information and entropy for random systems, we used numerical simulations. This analysis will show the quantitative functional working conditions of the remnant picture.

This paper is organized as follows. In SEC.~\ref{sec:pre}, we summarize two issues, where one is the brief introduction of the information loss problem and the other is the Page curve for bipartite systems. In SEC.~\ref{sec:inf}, we discuss on the information flow of tripartite systems, where we can demonstrate various scenarios. Finally, in SEC.~\ref{sec:dis}, we summarize important conclusions of this analysis and retrospect the implications for the information loss problem.

\section{\label{sec:pre}Information loss problem and information flow of bipartite systems}

In this section, we summarize two topics. First, we briefly summarize candidates of the resolution of the information loss problem. Second, we summarize and demonstrate the original Page curve for bipartite systems.

\subsection{Candidates of the resolution of the information loss problem}

\begin{figure}
\begin{center}
\includegraphics[scale=0.8]{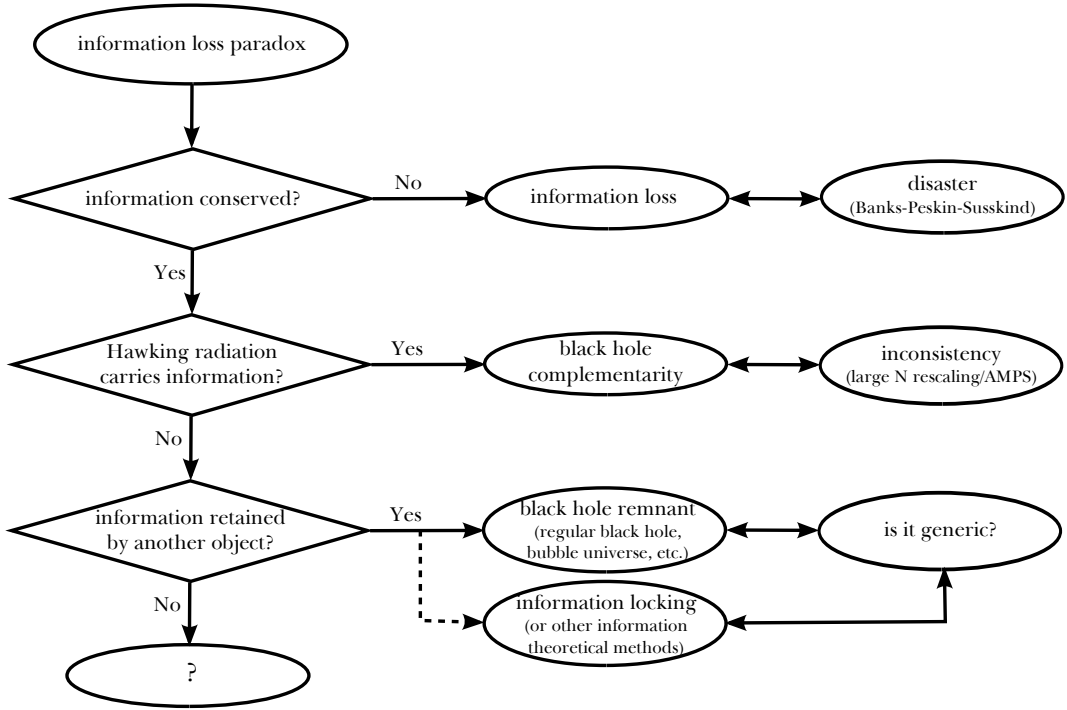}
\caption{\label{fig:flowchart}Flowchart on resolutions of the information loss problem \cite{Chen:2014jwq,Yeom:2016qec}.}
\end{center}
\end{figure}

In order to resolve the information loss problem, there have been lots of suggestions of the resolution (FIG.~\ref{fig:flowchart}). First of all, there has been an opinion that information inside a black hole would be disappeared \cite{Hawking:1976ra}. However, if this is the case, then there is a possibility that energy of the Universe may not be conserved, too \cite{Banks:1983by} (however, there are some authors who criticize \cite{Banks:1983by}, e.g., see \cite{Unruh:1995gn}). In any case, due to the AdS/CFT correspondence \cite{Maldacena:1997re}, many people could believe that information should be conserved at least for the anti-de Sitter background.

If information should be conserved, then the second question is this: \textit{what contains information} of the black hole? The first possibility is Hawking radiation. This idea was summarized by \textit{black hole complementarity} \cite{Susskind:1993if}, where the main assumptions are summarized as follows \cite{Ong:2016iwi}:
\begin{itemize}
\item[1.] \textit{Unitarity} from the black hole formation to its evaporation,
\item[2.] \textit{General relativity} for infalling observers (before touching the singularity),
\item[3.] \textit{Local quantum field theory} for asymptotic observers,
\item[4.] \textit{Statistical entropy} of the black hole is proportional to the \textit{event horizon area}, i.e., $\mathcal{A}/4 = \log n$, where $n$ is the number of states of the black hole and $\mathcal{A}$ is the area of the black hole horizon,
\item[5.] \textit{Existence of an observer} who can read and distinguish information from Hawking radiation.
\end{itemize}
If we accept these five assumptions, we can show inconsistency, either by the duplication experiment of information \cite{Yeom:2008qw} or by the Almheiri-Marolf-Polchinski-Sully (AMPS) thought experiment \cite{Almheiri:2012rt}.

If these five assumptions are not consistent, then we need to drop one of the assumptions. If we do not want to drop the first one (unitarity), then there are four possibilities.
\begin{itemize}
\item[P1.] AMPS thoughts that dropping general relativity for infalling observers is the most conservative opinion and they introduced so-called the firewall around the horizon where general relativity is broken \cite{Almheiri:2012rt}. One problem with this idea is that such a firewall is (if exists) possible to be naked to an asymptotic observer \cite{Hwang:2012nn}, and hence we need to drop not only general relativity but also the local quantum field theory for the asymptotic observer, too.
\item[P2.] If we drop the third assumption (local quantum field theory), then we may resolve the paradox. This can include a non-local information transfer \cite{Page:2013mqa}. This is a logical possibility, but we do not have a good mechanism to realize this idea yet. In this picture, we may say that (some part of) information is leaked gradually, where (some part of) information is not carried by typical Hawking radiation; some part of information can be attached to the form of the correction of Hawking radiation due to non-local effects.
\item[P3.] If we drop the fourth assumption, then there is no link between the black hole area and the number of states of the black hole. If a very small area object can store a huge number of states, then this can explain the information loss problem. This picture is briefly called by the remnant picture \cite{Chen:2014jwq}. In this picture, there are several variations.
\begin{itemize}
\item[P3-1.] The last Planck scale burst restores all information \cite{Ashtekar:2005cj} or the final Planck scale object carries all information \cite{Adler:2001vs}. For these models, we need to explain how can such a small object has a huge entropy. In addition, typically this picture has the infinite production problem \cite{Giddings:1994qt}. In addition, If there is no such a mechanism, then the final remnant can store some part of information but may not be all of them.
\item[P3-2.] In order to overcome this problem, one may imagine that information is stored by a huge interior object \cite{Christodoulou:2014yia} or a bubble universe \cite{Farhi:1989yr}. Regarding this, a typical criticism is that there is no generic principle or mechanism to justify large entropies.
\end{itemize}
Of course, there can be more variations from these possibilities (e.g., \cite{Horowitz:2003he,Smolin:2005tz}).
\item[P4.] The last possibility is that even though the total wave function is unitary, there may be no semi-classical observer who can see the restoration of information; only the superspace observer can read and distinguish information from the outcome of the black hole \cite{Maldacena:2001kr}. This picture can be called by the effective loss of information \cite{Sasaki:2014spa}, though this needs to be checked whether this can be a generic principle or not.
\end{itemize}

If Hawking radiation according to the local quantum field theory cannot carry information, while it should be carried by another object (i.e., P2 or P3), then these ideas can be approximately described not by the bipartite system but by the tripartite system, where the first part is the interior of the black hole (say $A$), the second part is the Hawking radiation from the black hole (according to the local quantum field theory, say $B$), and the third part is the `another' contribution (this can be anything, e.g., non-local contributions to typical Hawking radiation, the last burst, a Planck scale object, or a large interior inside the black hole, etc., say $C$). These `another' possibilities can be partly distinguished by the transferring process of states to $C$, whether a part of states are leaked gradually to $C$ (P2), a small part of states are transferred to $C$ only at the last stage of the black hole (P3-1), or almost all states are transferred to $C$ (P3-2). In the following sections, we investigate these hypotheses by using a tripartite toy-model system.

\subsection{\label{sec:pag}The Page curve for a bipartite system}

In this subsection, we consider a system with the total number of states $N$ \cite{Page:1993df,Page:1993wv}\footnote{Note that the total number of states may not be conserved; in this case, we can generalize our discussions \cite{Page:2013dx}, while in this paper we maintain the constant number of states for simplicity.}. We divide this system to two subsystems, where $A$ (inside of the black hole) has the number of states $n$ and $B$ (outside of the black hole) has the number of states $m$; $n$ and $m$ can vary as time goes on but $n \times m = N$ is fixed. Initially, we give $m=1$ and as time goes on, $n$ decreases and $m$ increases.

For a given state, we can define the density matrix:
\begin{eqnarray}
\rho = |\psi \rangle \langle \psi|.
\end{eqnarray}
Using this density matrix, we define the fine-grained entropy of $B$, or so-called the entanglement entropy of part $B$, by the formula:
\begin{eqnarray}
\rho_{B} &\equiv& \mathrm{tr}_{A} \rho,\\
S(B|A) &=& - \mathrm{tr} \rho_{B} \log \rho_{B}.
\end{eqnarray}
For the pure state, $S(B|A) = S(A|B)$.

Then we are ready to define information contents between inside and outside the black hole.
\begin{itemize}
\item[--] Information for $A$: $I_{A} \equiv \log n - S(A|B)$.
\item[--] Information for $B$: $I_{B} \equiv \log m - S(B|A)$.
\item[--] Mutual information between $A$ and $B$: $I_{AB} \equiv S(A|B) + S(B|A) - S(A \cup B)$. 
\end{itemize}
For $I_{A}$ and $I_{B}$, this definition of information can be a useful measure \cite{Lloyd:1988cn}, since it shows the difference between the exact thermal state and non-thermal states. In other words, if $I_{A}$ or $I_{B}$ is exactly zero, then there is no way to distinguish the state from the thermal equilibrium; on the other hand, if $I_{A}$ or $I_{B}$ is greater than zero, then this shows that $A$ or $B$ is biased from the thermal state, and hence there should be a way to read or distinguish information. One further interesting note is the case of the pure state. Then $S(A|B) = S(B|A)$ and $S(A\cup B) = 0$. Therefore, $I_{AB} = 2S(A|B)$ and $I_{A} + I_{B} + I_{AB} = \log N = \mathrm{const.}$ Hence, information will be conserved for all processes \cite{Alonso-Serrano:2015bcr}.

The problem is to calculate the entanglement entropy. We can further proceed by assuming that the system under consideration is pure and random \cite{Page:1993df}: if $1 \ll m \leq n$, then
\begin{eqnarray}
S(B|A) &=& \sum_{k=n+1}^{mn} \frac{1}{k} - \frac{m-1}{2n}\\
&\cong& \log m - \frac{m}{2n}.
\end{eqnarray}
Initially, the emitted information is $\cong m/2n$, and therefore is negligible. If $m > n$, since $S(B|A)=S(A|B)$ for a pure state, one gets
\begin{eqnarray}
S(B|A) &=& \sum_{k=m+1}^{mn} \frac{1}{k} - \frac{n-1}{2m}\\
&\cong& \log n - \frac{n}{2m}.
\end{eqnarray}
Thus, after $n$ becomes greater than $m$, the information emitted is given by $\cong \log m - \log n + n/2m$, and then it gradually increases. The time of $n \sim m$ is called by the Page time, and from this time, we can distinguish information from radiation.

\begin{figure}
\begin{center}
\includegraphics[scale=0.4]{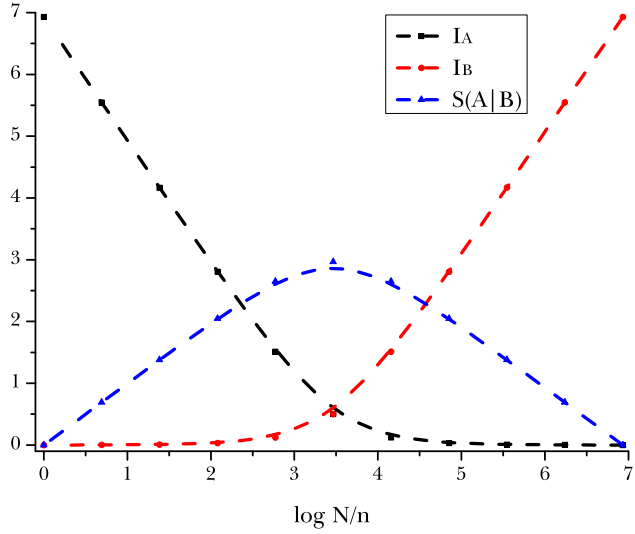}
\caption{\label{fig:bipart}Information flow, where $I_{A}$ (black), $I_{B}$ (red), and $S(A|B)$ (blue).}
\end{center}
\end{figure}

For a toy model of the spin-$1/2$ system, the Page curve is produced by the following manner. Here, we introduce $J$ number of spins. For each spin, we assign coefficients:
\begin{eqnarray}
|\psi_{i} \rangle = c_{i1} \left( \begin{array}{c}
                            1 \\
                            0
                          \end{array} \right)
+ c_{i2} \left( \begin{array}{c}
                            0 \\
                            1
                          \end{array} \right),
\end{eqnarray}
where $1 \leq i \leq J$ is a spin index and $|c_{i1}|^{2} + |c_{i2}|^{2} = 1$ with complex values $c_{i1}$ and $c_{i2}$. The total state for the non-entangled case is
\begin{eqnarray}
|\psi \rangle = \bigotimes_{i=1}^{J} |\psi_{i} \rangle.
\end{eqnarray}
However, in general, the states can be entangled each other. For simplicity and generality, we assume \textit{random} mixing between each particles. For example, for two spin-1/2 particles,
\begin{eqnarray}
|\psi_{A} \rangle &=& \sum_{i=1}^{2} c^{A}_{i} |i \rangle_{A},\\
|\psi_{B} \rangle &=& \sum_{j=1}^{2} c^{B}_{j} |j \rangle_{B},\\
|\psi_{AB} \rangle &=& \sum_{ij} c_{ij} |ij \rangle,
\end{eqnarray}
while $|ij \rangle = |i \rangle_{A} \bigotimes |j \rangle_{B}$, $| 1 \rangle = \left( \begin{array}{c}
                            1 \\
                            0
                          \end{array} \right)$, and
$| 2 \rangle = \left( \begin{array}{c}
                            0 \\
                            1
                          \end{array} \right)$, then
\begin{eqnarray}
c_{ij} = d_{ij} c^{A}_{i}c^{B}_{j},
\end{eqnarray}
where we introduce an arbitrary mixing factor $d_{ij}$. If $|d_{ij}|=1$, then all states are separable; otherwise, it is not separable and is entangled \cite{LeBellac}.

Based upon this quantum state, we calculate the density matrix as we successively trace out particles. 
Initially the density matrix is a $(2^{J} \times 2^{J})$-matrix
\begin{equation}
\rho_J = | \psi \rangle \langle \psi |
\end{equation}
which can be interpreted as a $(2 \times 2)$ block matrix whose elements are $(2^{J-1} \times 2^{J-1})$ matrices. 
As we trace out the first particle by summing the two block diagonal components of $\rho_J$, the element of $a$-th row and $b$-th column of the reduced density matrix $\rho_{J -1}[a,b]$ becomes 
\begin{equation}
\rho_{J -1}[a,b]= \rho_J[a,b] + \rho_J[a+2^{J-1},b+2^{J-1}].
\end{equation}
If we continue to repeat this process, we can describe the situation that particles move from part $A$ to part $B$. A numerical demonstration of this process is in FIG.~\ref{fig:bipart}.

\section{\label{sec:inf}Information flow of tripartite system}

\subsection{Information contents for a tripartite system}

Now, we consider a tripartite system that demonstrates various ideas of the information loss problem. We divide by three parts: interior of a black hole ($A$ with the number of states $n$), exterior of a black hole (or typical Hawking radiation, $B$ with the number of states $m$), and another part that can be interpreted in various ways ($C$ with the number of states $\ell$). For simplicity, we give the condition that the total number of states is a fixed constant $N$: $n \times m \times \ell = N$. In the initial condition $n = N$ and $m = \ell = 1$; and in the final condition $n = 1$, while $m$ or $\ell$ depends on scenarios. For spin-1/2 particles, the number of states is related to the number of particles: as the total number of particles is $n_{\mathrm{tot}}$, the relation is $N = 2^{n_{\mathrm{tot}}}$, $n = 2^{n_{A}}$, $m = 2^{n_{B}}$, and $\ell = 2^{n_{C}}$ with $n_{\mathrm{tot}} = n_{A} + n_{B} + n_{C}$, where $n_{A,B,C}$ are number of particles for each part $A$, $B$, and $C$, respectively.

Then we can easily generalize the information contents of a tripartite system \cite{Rota:2015wge}. First, information for each sectors are
\begin{itemize}
\item[--] Information for $A$: $I_{A} \equiv \log n - S(A|B \cup C)$.
\item[--] Information for $B$: $I_{B} \equiv \log m - S(B|A \cup C)$.
\item[--] Information for $C$: $I_{C} \equiv \log \ell - S(C|A \cup B)$.
\end{itemize}
Mutual information between $AB$, $BC$, and $AC$ are as follows:
\begin{itemize}
\item[--] Mutual information between $A$ and $B$: $I_{AB} \equiv S(A|B \cup C) + S(B|A \cup C) - S(A \cup B | C)$. 
\item[--] Mutual information between $B$ and $C$: $I_{BC} \equiv S(B|C \cup A) + S(C|B \cup A) - S(B \cup C | A)$.
\item[--] Mutual information between $A$ and $C$: $I_{AC} \equiv S(A|C \cup B) + S(C|A \cup B) - S(A \cup C | B)$.
\end{itemize}
Finally, we define the mutual information for $ABC$:
\begin{itemize}
\item[--] Tripartite information: $I_{ABC} \equiv S(A|B \cup C) + S(B|A \cup C) + S(C|A \cup B) - S(A\cup B | C) - S(A \cup C | B) - S(B \cup C | A) + S(A \cup B \cup C)$. 
\end{itemize}

For a pure state, $S(X|Y) = S(Y|X)$ for any $X$ and $Y$ as well as $S(A \cup B \cup C) = 0$. Therefore, these definitions can be simplified as follows:
\begin{itemize}
\item[--] $I_{A} = \log n - S(A|B \cup C)$.
\item[--] $I_{B} = \log m - S(B|A \cup C)$.
\item[--] $I_{C} = \log \ell - S(C|A \cup B)$.
\item[--] $I_{AB} = S(A|B \cup C) + S(B|A \cup C) - S(C|A \cup B)$. 
\item[--] $I_{BC} = S(B|C \cup A) + S(C|B \cup A) - S(A|B \cup C)$.
\item[--] $I_{AC} = S(A|C \cup B) + S(C|A \cup B) - S(B|A \cup C)$.
\item[--] $I_{ABC} = 0$.
\end{itemize}
For this pure state, $I_{A} + I_{B} + I_{C} + I_{AB} + I_{AC} + I_{BC} = \log N = \mathrm{const.}$ and hence the total information is conserved.

\subsection{Numerical setup}

In order to generalize the Page curve for tripartite systems using numerical techniques, we use a toy model of spins. This was successfully and completely reproduced the Page curve for bipartite systems (SEC.~\ref{sec:pag}). Following this, we can expect that the toy model of spins will reproduce essential properties of tripartite systems.

First, we consider the total system as a tensor product of three subsystems $A$ (inside of the black hole), $B$ (Hawking radiation), and $C$ (another part) in the Hilbert space: $A\otimes B\otimes C$. We know that the number of states of $A$ is $N$ and of $B$ and $C$ are $1$ in the beginning. We suppose that one particle is released from $A$ to $B$ or $C$ at each step. Then as particle comes out from $A$, the number of states of $A$ decreases and that of $B$ or $C$ increases since the total number of states is a constant. As the number of states of each subsystem changes, reduced density matrices also become different. We calculate the entanglement entropy, the information, and the mutual information from the reduced density matrix of each system at each step.

The number of states of the total system is $N=n \times m\times \ell$. Hence, the quantum states of particles in total system have the form
\begin{eqnarray}
|\psi\rangle=\sum_{i_{1}, i_{2}, ..., i_{n_{\mathrm{tot}}}}c_{i_{1}, i_{2}, ..., i_{n_{\mathrm{tot}}}}|i_{1}, i_{2}, ..., i_{n_{\mathrm{tot}}} \rangle,
\end{eqnarray}
where $|i_{1}, i_{2}, ..., i_{n_{\mathrm{tot}}}\rangle$ represents the spin states of particles. In order to make arbitrary quantum states, we set ${c_{i_{1}, i_{2}, ..., i_{n_{\mathrm{tot}}}}}$ as a random number. Then the density matrix is written as $\rho=|\psi\rangle\langle\psi|$. 

In order to calculate each system's information, we operate partial traces to the density matrix. We do partial trace for the system $A$ first and do for $B$ and $C$ next. 
The entries of $a$-th row and $b$-th column of the reduced density matrices are obtained as follows: 
\begin{eqnarray}
\rho_{A}[i,j] &=& \sum_{k=1}^{m\ell}{\rho[n(i-1)+k, n(j-1)+k]},\label{eq:1}\\
\rho_{B \cup C}[i,j]&=&\sum_{k=1}^{n}{\rho[m\ell(k-1)+i, m\ell(k-1)+j]},\label{eq:2}\\
\rho_{B}[i,j]&=&\sum_{k=1}^{\ell}{\rho_{B\cup C}[m(i-1)+k,m(j-1)+k]},\label{eq:3}\\
\rho_{C}[i,j]&=&\sum_{k=1}^{m}{\rho_{B\cup C}[\ell(k-1)+i,\ell(k-1)+j]}\label{eq:4}.
\end{eqnarray}
We explain how to get Eq.~(\ref{eq:1}) from the entire density matrix $\rho$ which is an $(nml \times nml)$-matrix. 
In order to obtain $\rho_{A}$, we interpret $\rho$ as an $(ml \times ml)$-block matrix whose component is an $(n \times n)$-matrix\footnote{Hence, for $\rho_{A}[i,j]$, $i$ and $j$ denote the orthonormal basis of the system $A$, where there are $2^{n_{\mathrm{A}}}$ different states and hence $1 \leq i,j \leq 2^{n_{A}}$. For the other cases, e.g., $\rho_{B\cup C}[i,j]$, $\rho_{B}[i,j]$, and $\rho_{C}[i,j]$, $i$ and $j$ denote the orthonormal basis of the given system, e.g., $B \cup C$, $B$, and $C$, respectively.
Then we obtain $\rho_A$ by tracing out $B$ and $C$, which means that we sum up $(i, j)$-th components of $(n \times n)$-matrices from the block diagonals of $(ml \times ml)$-matrix. 
In order to obtain Eq.~(\ref{eq:2}) from $\rho$, we interpret $\rho$ as an $(n \times n)$-matrix whose component is an $(ml \times ml)$-matrix. Then tracing out $A$ means that we sum up $(i, j)$-th components of $(ml \times ml)$-matrices from the block diagonals of  $(n \times n)$-matrix. In this way, we can understand how to trace-out systems. Consequently, we get three reduced density matrices of $A$, $B$, and $C$. Now, information for $A$ is $\log n - S(A|B \cup C)$. Here, the entanglement entropy of $A$ is written as follows:
\begin{eqnarray}
S(A|B \cup C) = -\mathrm{tr} \rho_{A} \log\rho_{A}.
\end{eqnarray}
The other parts of information can be defined in a similar way. }

Due to the limitation of the computing power, in this paper, we use $10$ spin-$1/2$ particles, so that the total number of states is $N = 2^{10}$. Since we are using random variables, in order to make statistically relevant results, we repeated fifty times the same numerical experiments and averaged the repeated results for information quantities\footnote{Note that as the number of particles increases (ten particles are way enough), the standard deviation becomes negligibly small and hence we do not plot them in the following figures.}. In order to show the physical tendency, we smoothly connected by B-splines. Finally, we plotted graphs of information of subsystems $A$, $B$, and $C$ and various mutual information. The parameter of $x$-axis is $\log(N/n)$ to represent the number of spins emitting from $A$ that is a monotone function of the time.

\begin{figure}
\begin{center}
\includegraphics[scale=0.25]{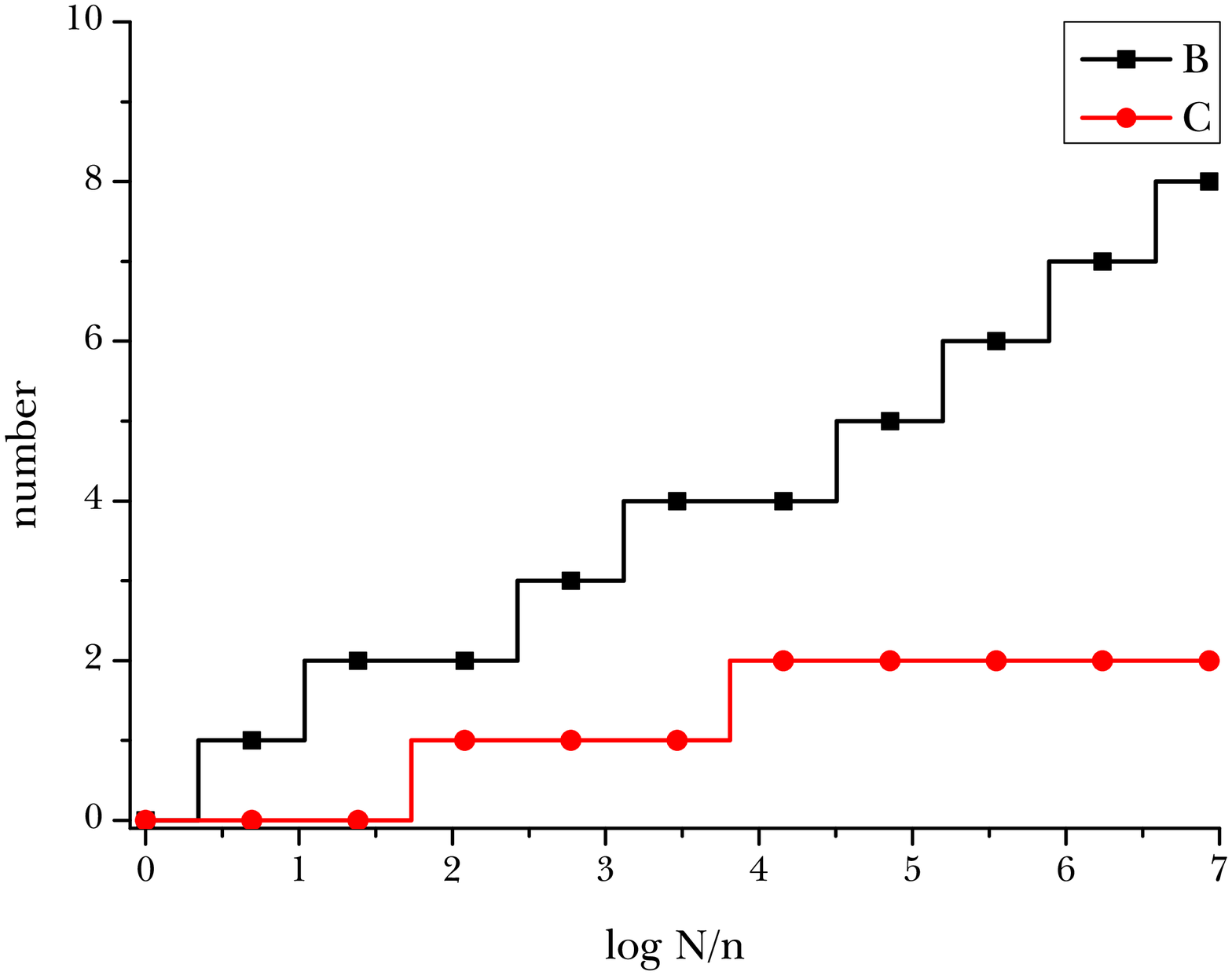}
\includegraphics[scale=0.25]{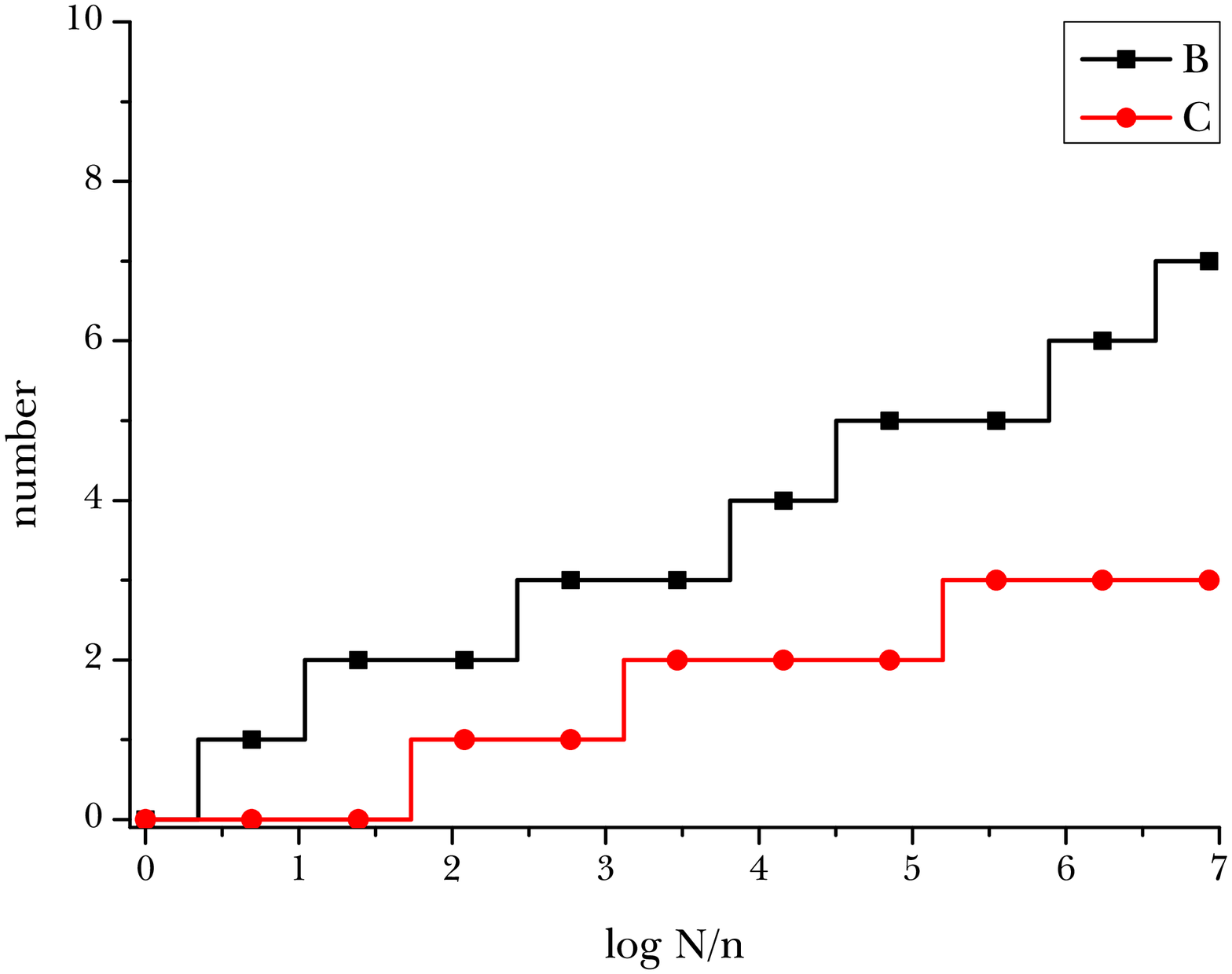}
\caption{\label{fig:P2_12}Number of particles transferred to $B$ and $C$ for FIG.~\ref{fig:P2_82} (left) and FIG.~\ref{fig:P2_73} (right).}
\includegraphics[scale=0.25]{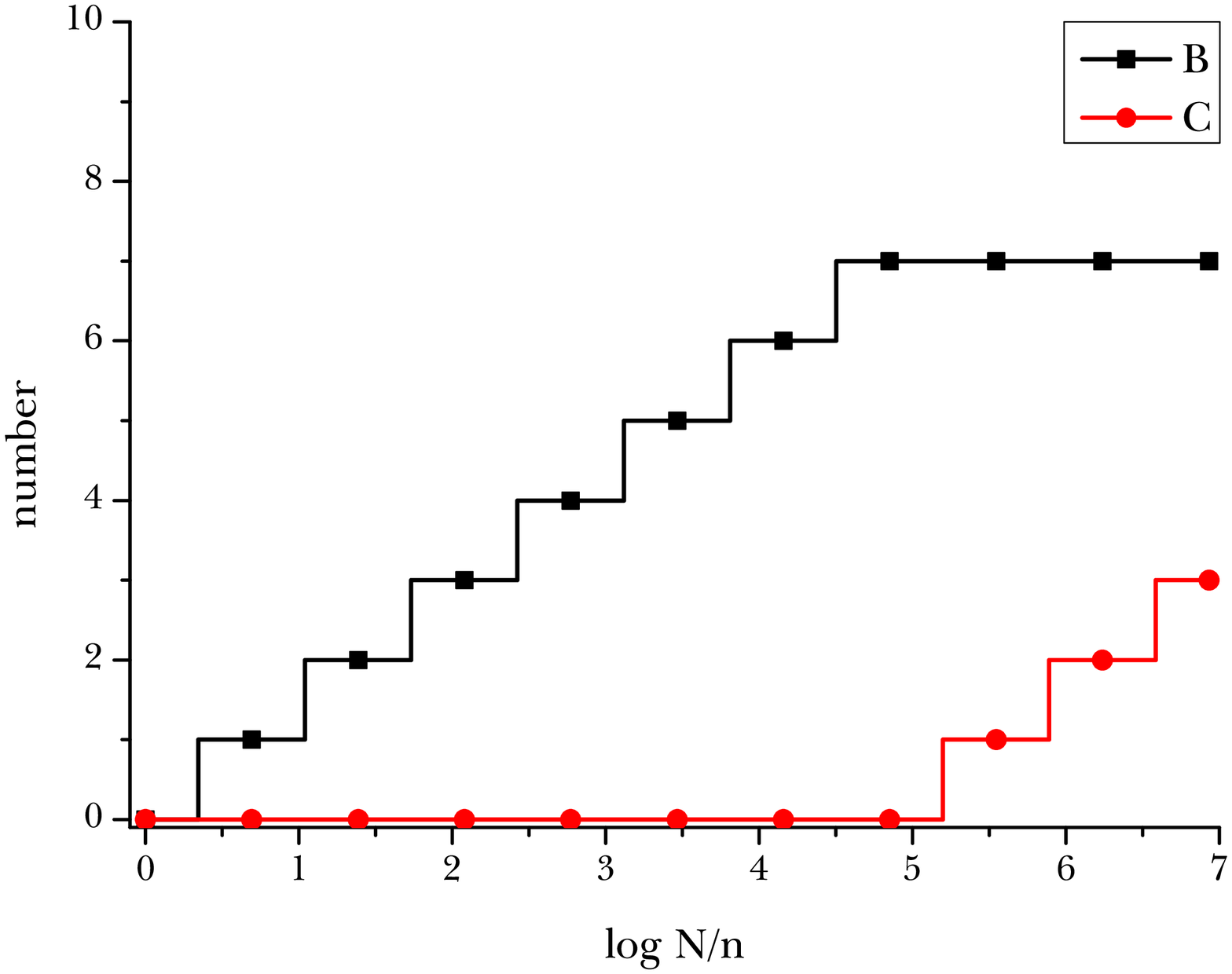}
\includegraphics[scale=0.25]{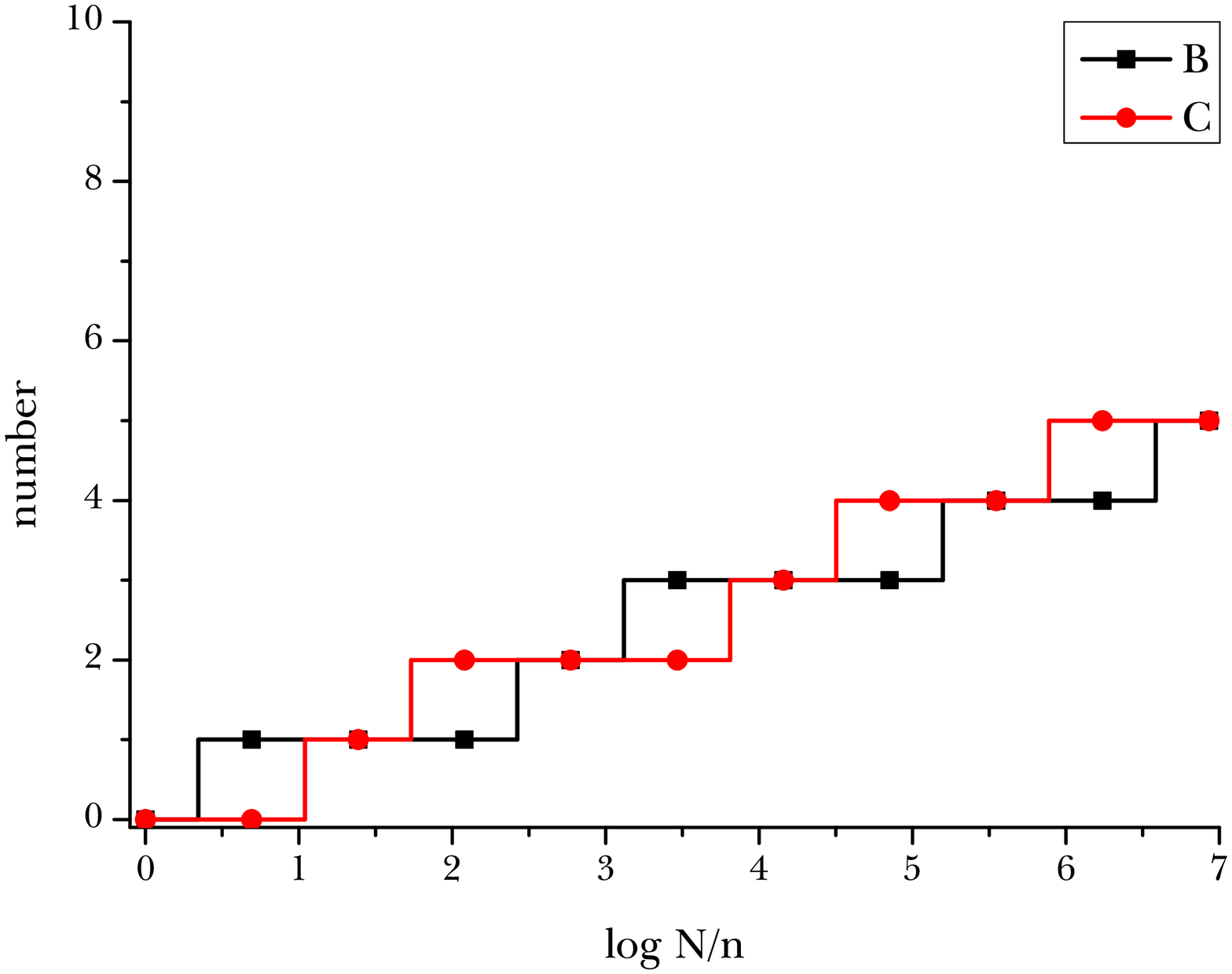}
\caption{\label{fig:P31_12}Number of particles transferred to $B$ and $C$ for FIG.~\ref{fig:P31_73} (left) and FIG.~\ref{fig:P31_55} (right).}
\includegraphics[scale=0.25]{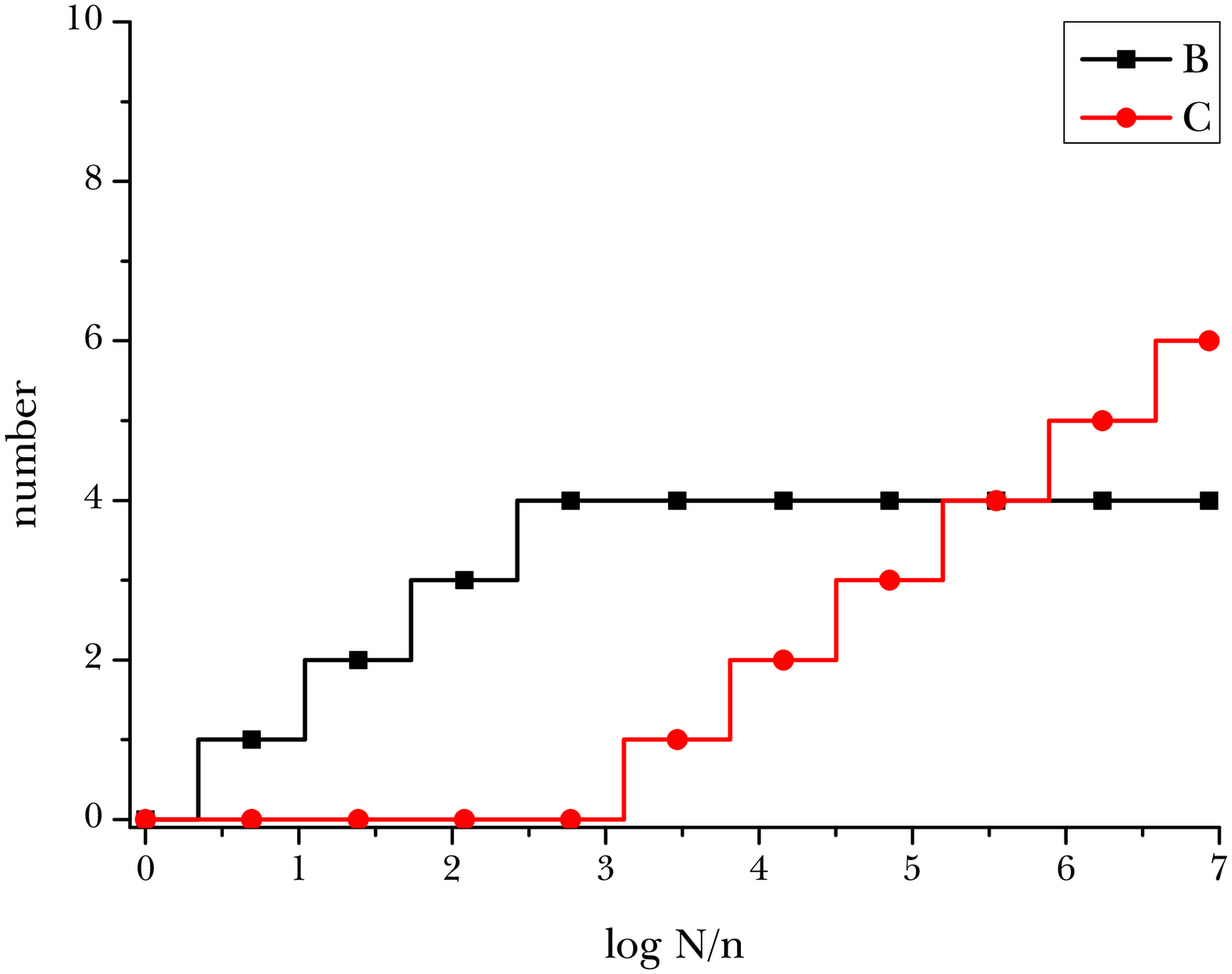}
\includegraphics[scale=0.25]{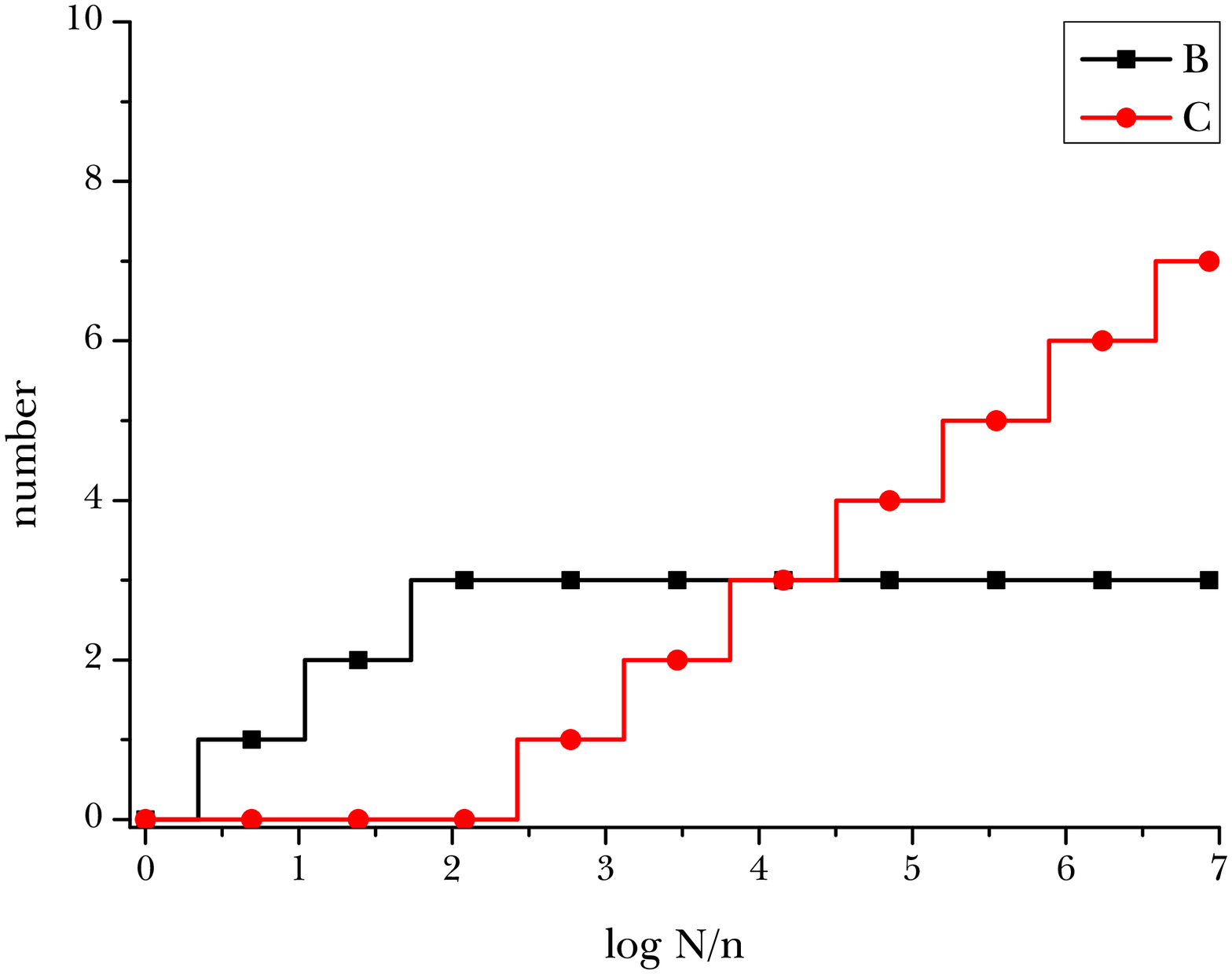}
\caption{\label{fig:P32_12}Number of particles transferred to $B$ and $C$ for FIG.~\ref{fig:P32_46} (left) and FIG.~\ref{fig:P32_37} (right).}
\end{center}
\end{figure}

\subsection{Various scenarios}

In tripartite systems, there is no unique way to move states from $A$ to $B$ or $C$. This depends on the detailed scenarios. As we take particles from $A$ to either $B$ or $C$, we assign the history of subsystem $C$, i.e., $\ell(t)$ (e.g., see FIGs.~\ref{fig:P2_12}, \ref{fig:P31_12}, and \ref{fig:P32_12}), that mimics a certain scenario (here, $t$ can be simply chosen by $\log N/n$). Note that this $t$ does not necessarily have an explicit meaning at this stage, but just simply implies the time ordering of the evolution\footnote{If we give a physical condition for the time, e.g., if we assume that the entropy is proportional to the black hole area, then we can specify more details of the time evolution. However, we will not assume such details, and hence, there is an ambiguity to define the dynamics of the time evolution. This can be a strong point of this approach, since what we are investigating does not necessarily depend on the semi-classical or quantum gravitational details, but only depend on number of states of each part.}. Also, we can choose $m_{\mathrm{fin}}$ or $\ell_{\mathrm{fin}}$, where this is the number of states at the last stage. By varying the history as well as the number of states at the last stage, we can choose the following variations of scenarios.
\begin{itemize}
\item[P2:] $m_{\mathrm{fin}} \gg \ell_{\mathrm{fin}}$ and $\ell$ gradually increases. This mimics the scenario that (some, but not all of) information is leaked consecutively from the black hole, either by non-local effects or bubble universes.
\item[P3-1:] $m_{\mathrm{fin}} \gg \ell_{\mathrm{fin}}$ and $\ell$ increases only at the last stage. This mimics the scenario that the last stage of the black hole (either by a remnant or the last burst) contains some part of information.
\item[P3-2:] $m_{\mathrm{fin}} \lesssim \ell_{\mathrm{fin}}$ and $\ell$ gradually increases. This mimics the scenario that almost all information is stored in the internal space (e.g., a bubble universe) or is actively transferred to somewhere by unknown (e.g., non-local) effects.
\end{itemize}
Of course, these three categories are rather heuristic and they require more clarifications for real applications. For more detailed histories of particle transferring processes, we illustrate in FIGs.~\ref{fig:P2_12}, \ref{fig:P31_12}, and \ref{fig:P32_12}.

\begin{figure}
\begin{center}
\includegraphics[scale=0.29]{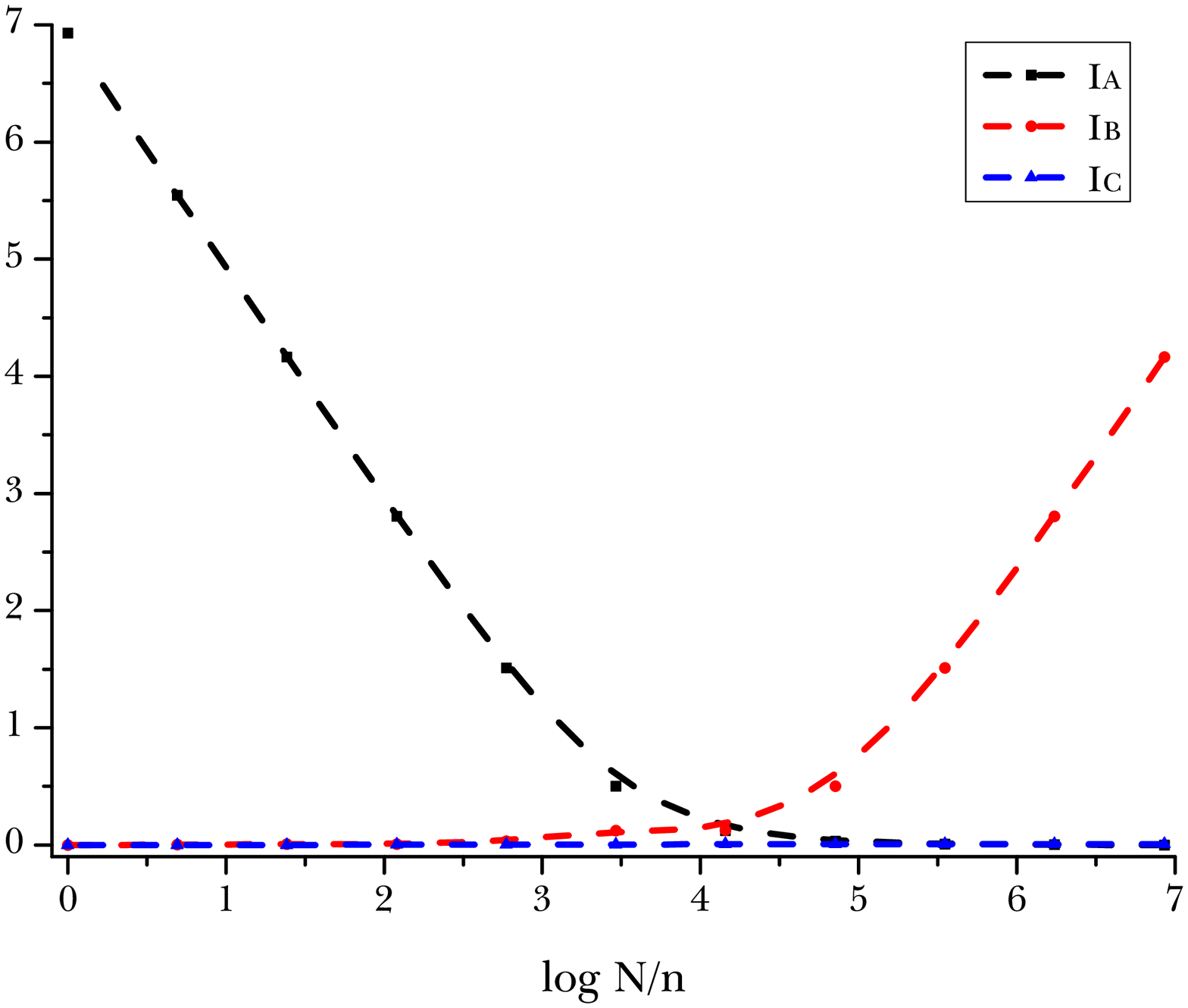}
\includegraphics[scale=0.29]{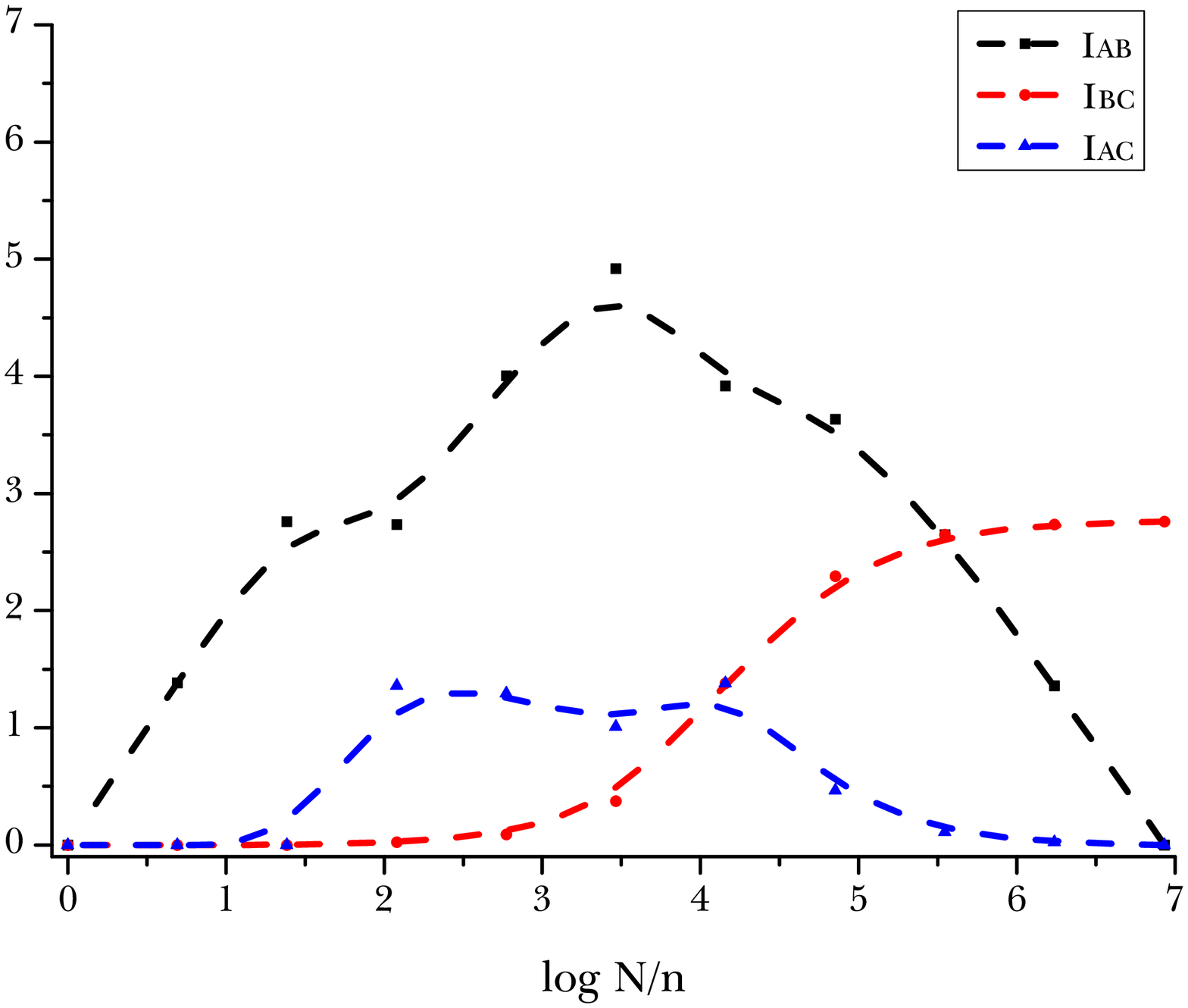}
\caption{\label{fig:P2_82}Information (left) and mutual information (right) flow for P2 (with $m_{\mathrm{fin}} = 2^{8}$ and $\ell_{\mathrm{fin}} = 2^{2}$).}
\includegraphics[scale=0.29]{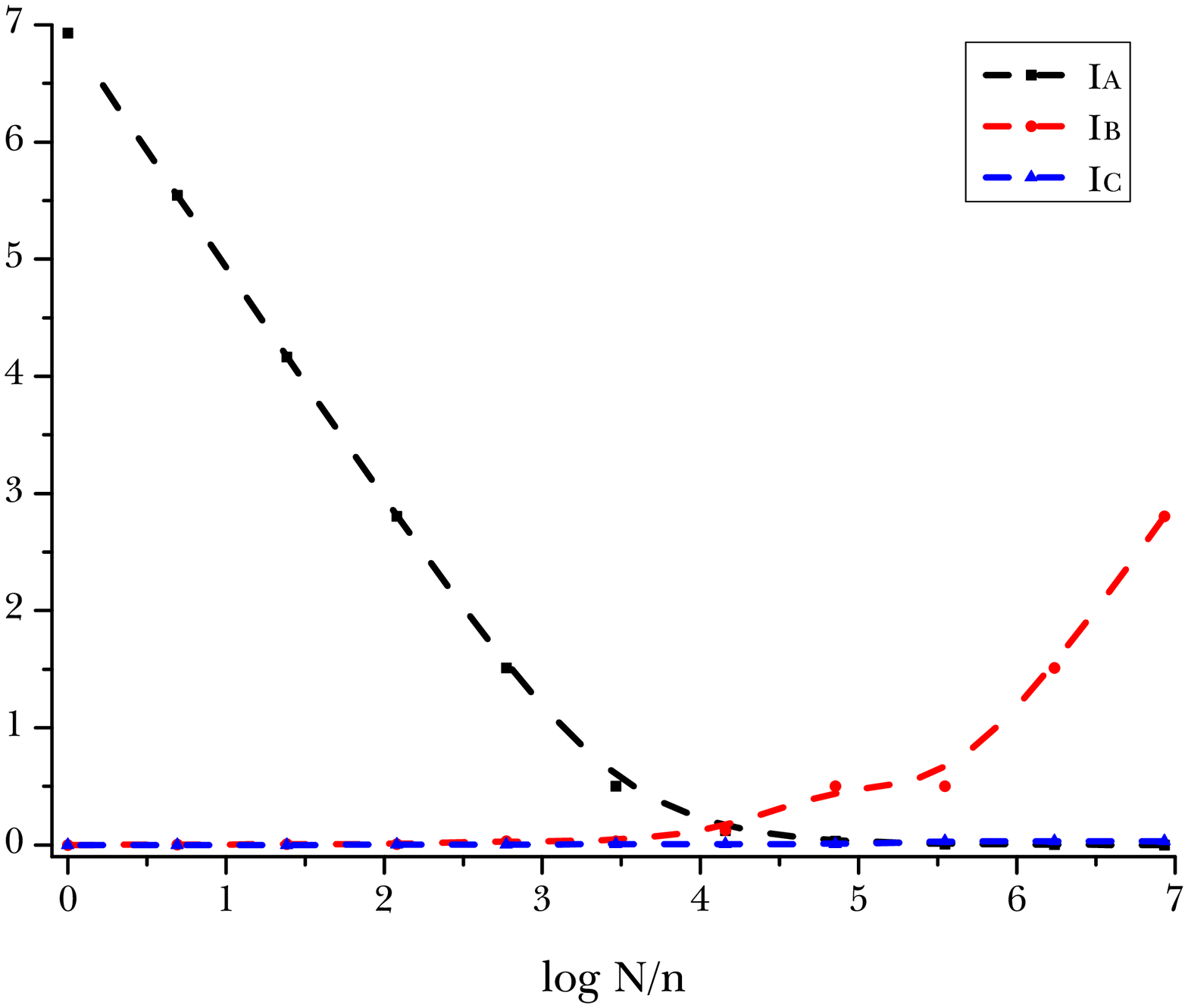}
\includegraphics[scale=0.29]{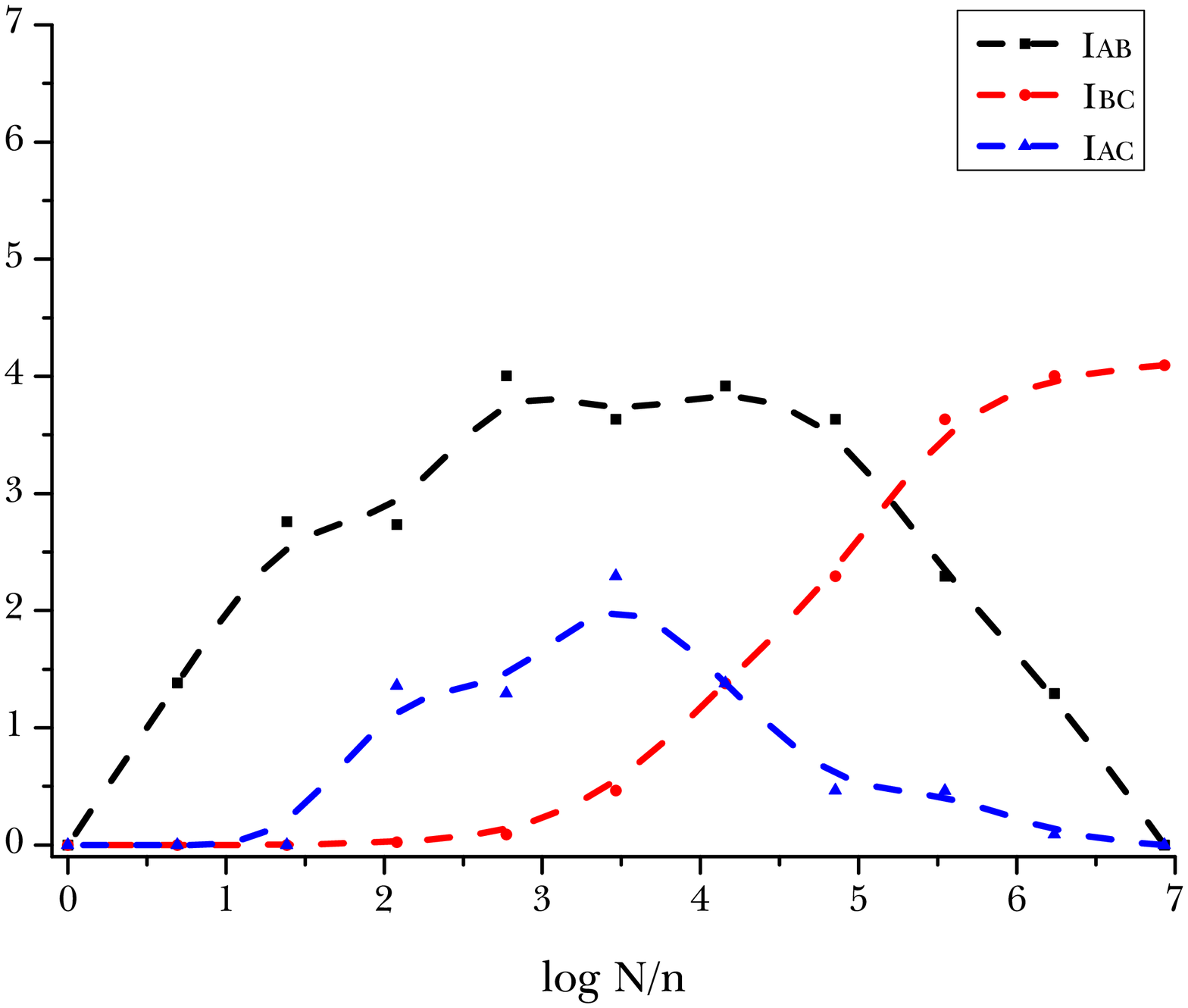}
\caption{\label{fig:P2_73}Information (left) and mutual information (right) flow for P2 (with $m_{\mathrm{fin}} = 2^{7}$ and $\ell_{\mathrm{fin}} = 2^{3}$).}
\end{center}
\end{figure}

\subsubsection{P2: Consecutive information leak}

In the scenario P2, we assume $m_{\mathrm{fin}} \gg \ell_{\mathrm{fin}}$ and $\ell$ gradually increases as demonstrated in FIGs.~\ref{fig:P2_82} and \ref{fig:P2_73}. Since $\ell$ gradually increases, the mutual information during the evaporation can be quite complicated. Although the mutual information $I_{AB}$ and $I_{AC}$ wiggly behave (probably, this is due to the fact that our number of particles are too small), there is a clear tendency that they start from zero, reach the maximum around the halfway, and finally become zero. In addition, similar as the case of the bipartite system, information of Hawking radiation $I_{B}$ begins to increase around the halfway point.

Although the intermediate information transfer by mutual information is quite complicated, the information should be transferred by two channels at the last stage: the information of $B$, i.e., $I_{B}$, and the mutual information between $B$ and $C$, i.e., $I_{BC}$. Therefore, the existence of $C$ (probably, a non-local transfer of information) drastically changes the type of information. In other words, finally, Hawking radiation should contain a certain amount of information, but it cannot contain all of them; the other part of information should be the form of the mutual information between Hawking radiation and its corrections (denoted by $C$).

One more interesting remark is that $I_{C}$ is negligible all over the time. Therefore, can this mean that such a non-local effect does not necessarily mean a causality violation, but just a well-known non-locality of quantum mechanics, such as Einstein-Podolski-Rosen pairs \cite{EPR}? We cannot decide this question now, but surely \textit{(even if there is a non-local transfer of information) an observer in $C$ cannot see any transferred information from $A$}. In any case, even though there is a gradual information leak, if the number of states is smaller than that of Hawking radiation, then it is possible that the part $C$ can do a role of information storage (in the form of the mutual information) even though $C$ itself does not contain information.

\subsubsection{P3-1: Remnants, or the last burst}

\begin{figure}
\begin{center}
\includegraphics[scale=0.29]{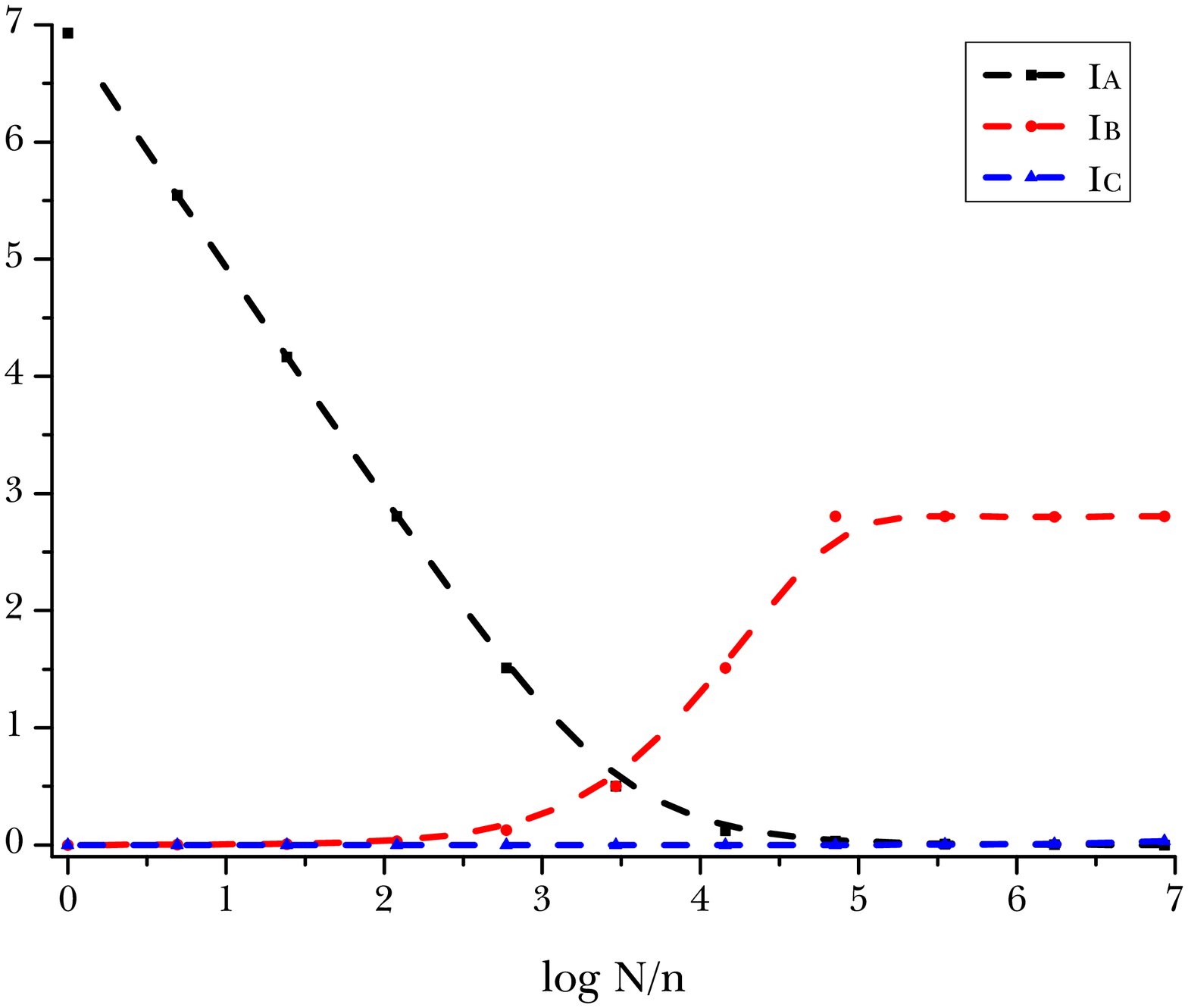}
\includegraphics[scale=0.29]{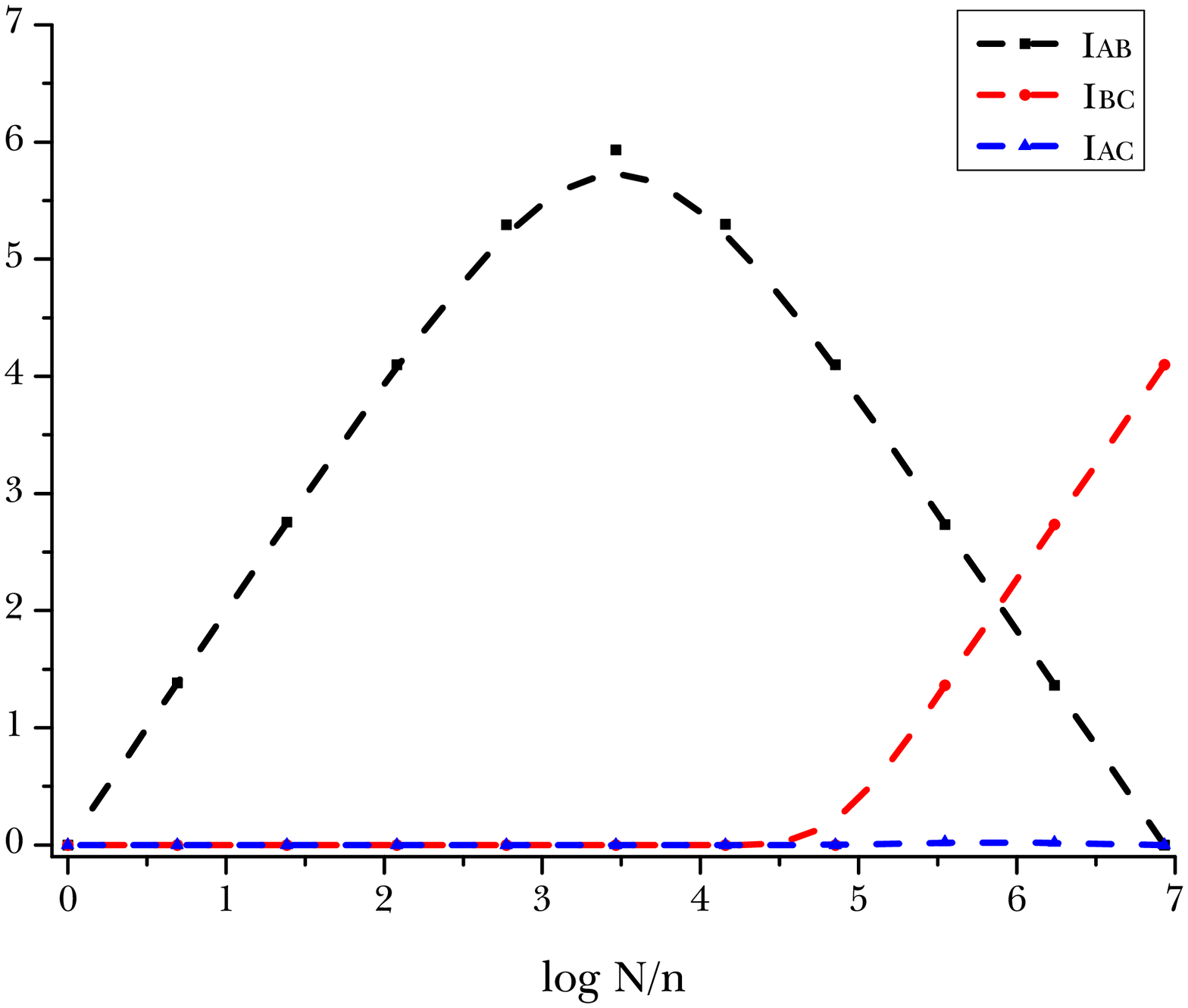}
\caption{\label{fig:P31_73}Information (left) and mutual information (right) flow for P3-1 (with $m_{\mathrm{fin}} = 2^{7}$ and $\ell_{\mathrm{fin}} = 2^{3}$).}
\includegraphics[scale=0.29]{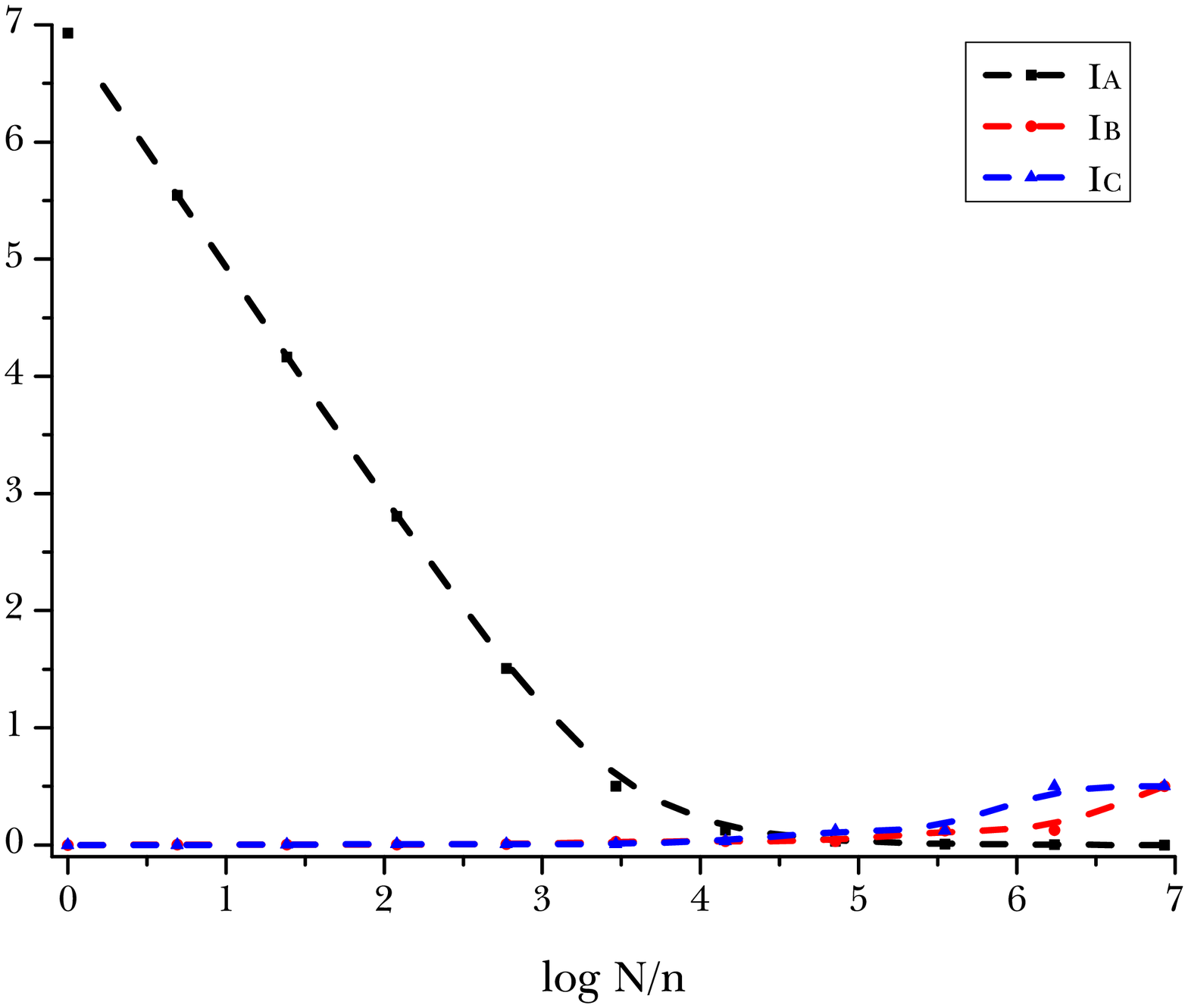}
\includegraphics[scale=0.29]{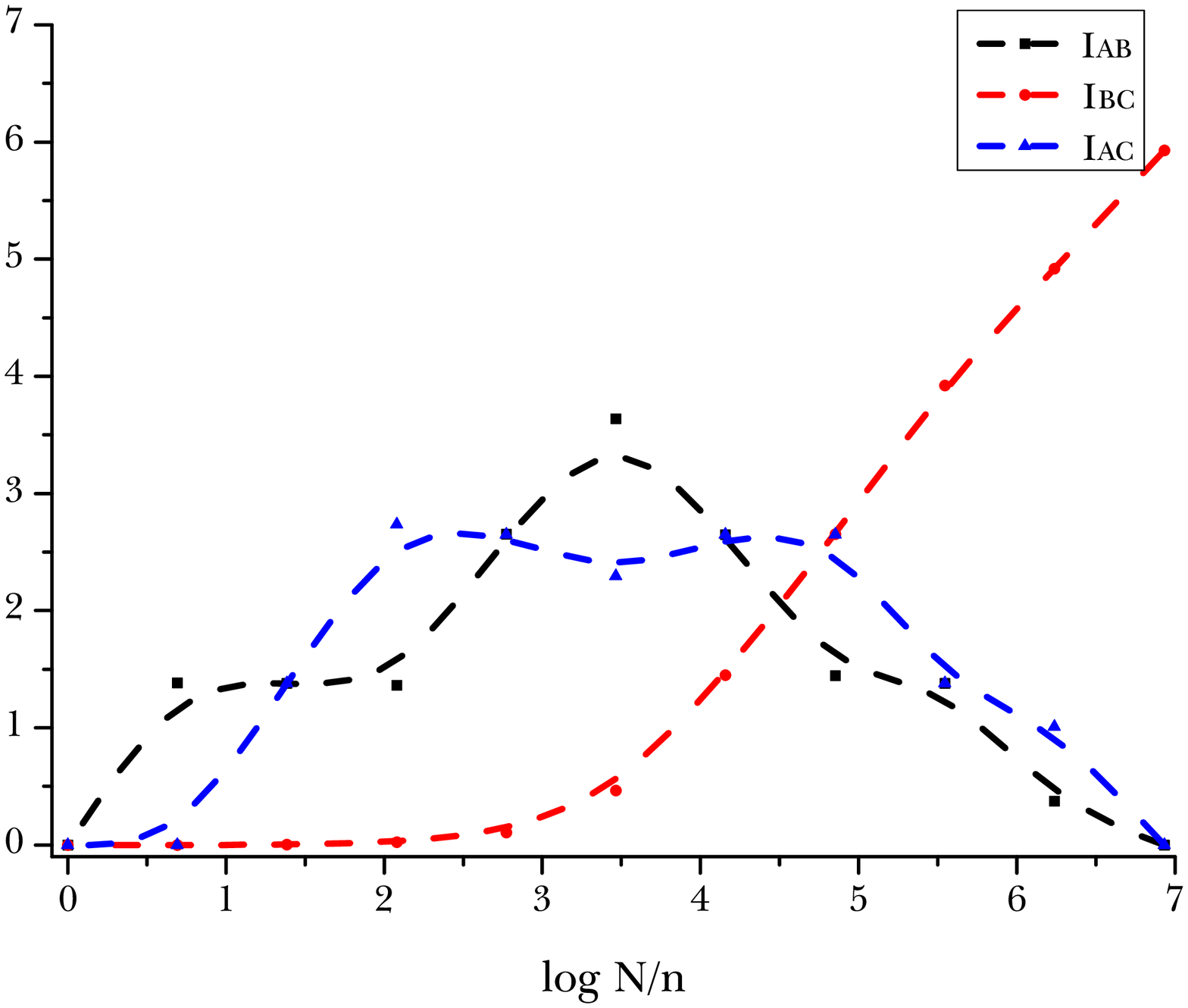}
\caption{\label{fig:P31_55}Information (left) and mutual information (right) flow for $m_{\mathrm{fin}} = 2^{5}$ and $\ell_{\mathrm{fin}} = 2^{5}$.}
\end{center}
\end{figure}

In the scenario P3-1, we also assume $m_{\mathrm{fin}} \gg \ell_{\mathrm{fin}}$, but the intermediate history is different compared to P2. In this case, particles move to $C$ only at the last stage of the evolution. FIG.~\ref{fig:P31_73} demonstrates this case, where this last stage can have a very long life time (usual remnants) or a very short life time (e.g., the last burst) in terms of physical time scales.

In FIG.~\ref{fig:P31_73}, around the time $\log N/n \sim 5$, particles starts to move to $C$. Before this time, the curve well coincides with that of the bipartite system. However, after that time, $I_{B}$ stops to increase and the information is transferred to the form of the mutual information $I_{BC}$.

The pattern of information distributions at the intermediate step is significantly different from that of P2, though similar at the last step (FIG.~\ref{fig:P2_73}). In this sense, by measuring some information contents, e.g., the entanglement entropy of part $B$, one may distinguish the process whether there is a continuous leak of information or not.

\begin{figure}
\begin{center}
\includegraphics[scale=0.29]{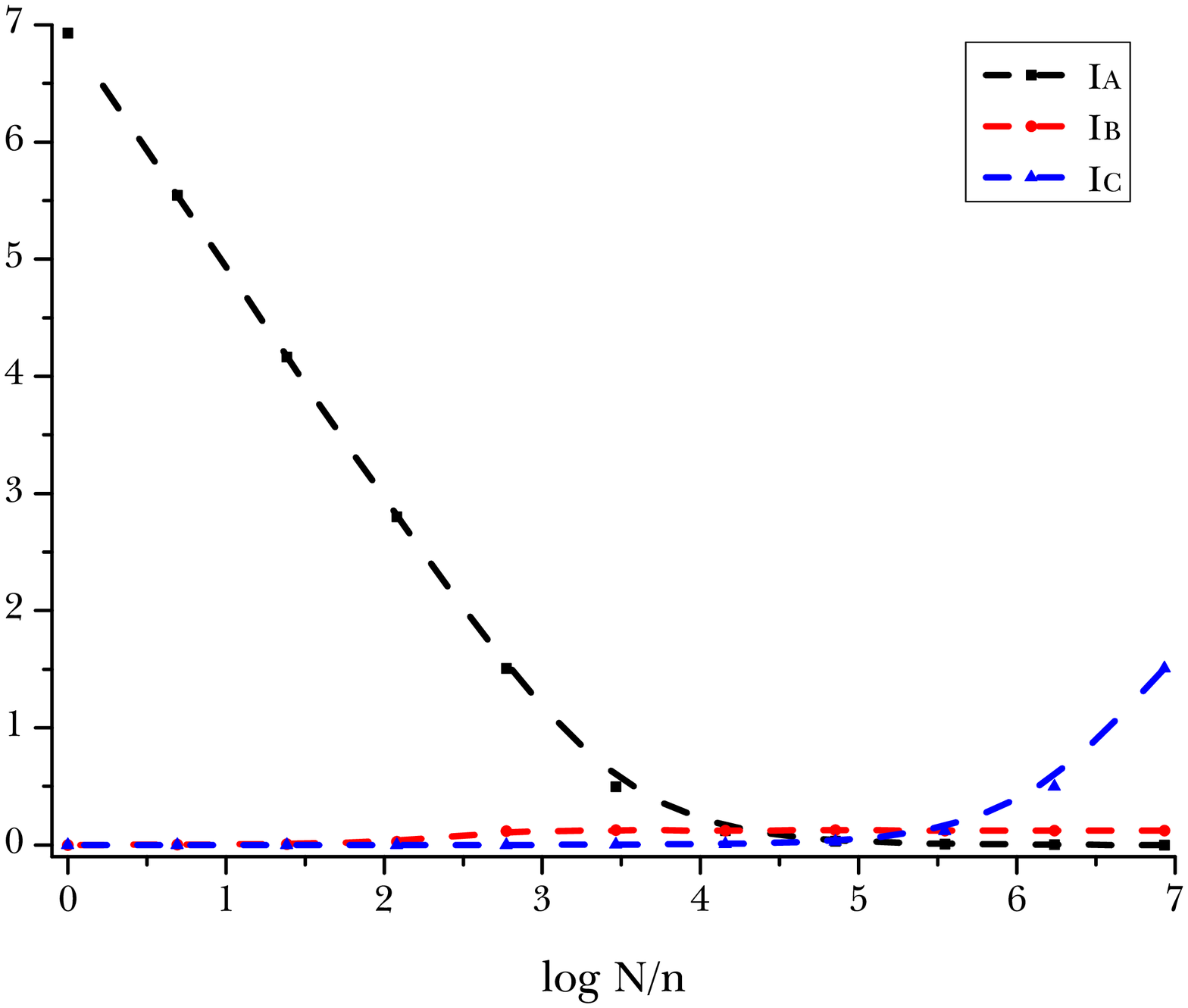}
\includegraphics[scale=0.29]{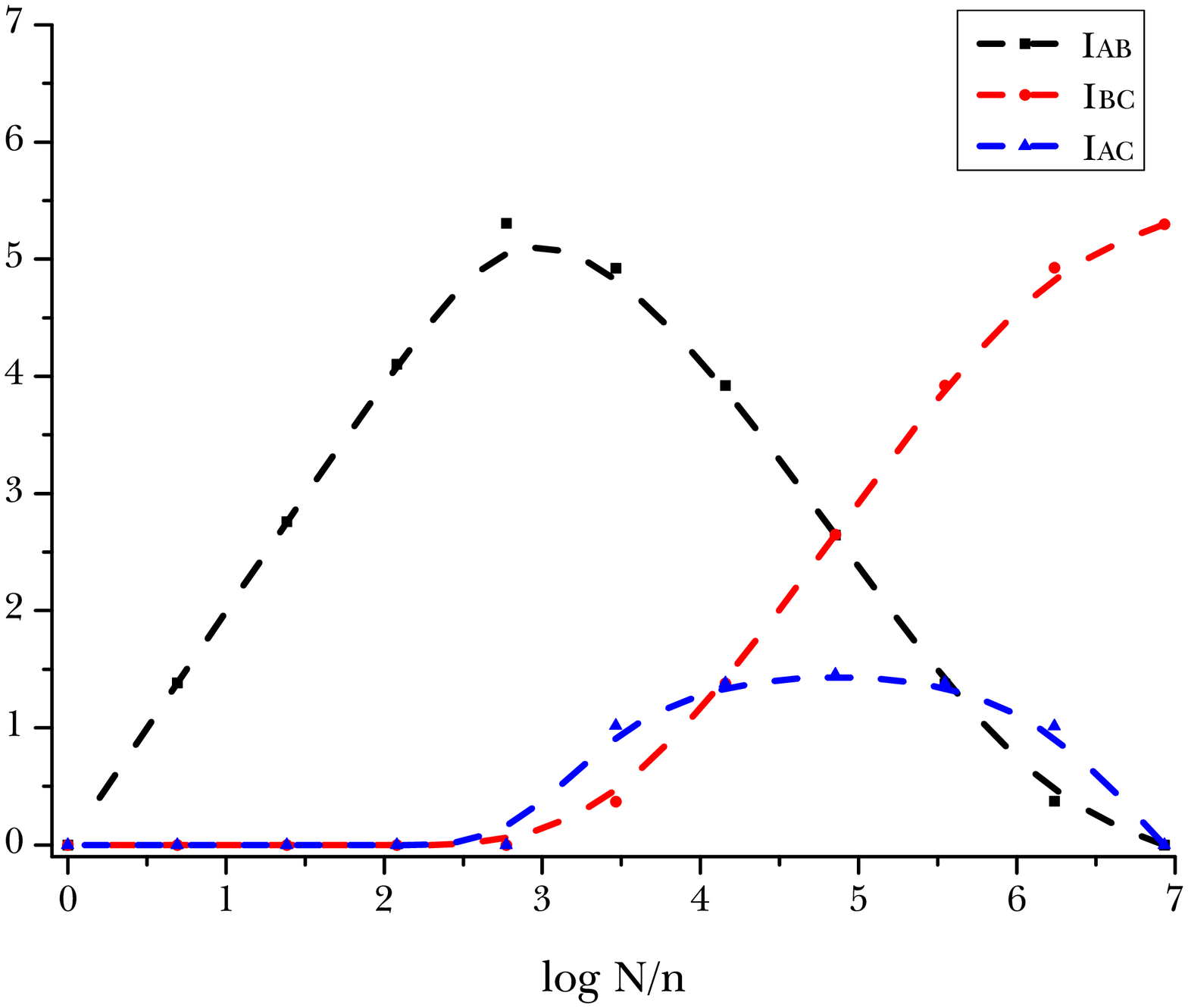}
\caption{\label{fig:P32_46}Information (left) and mutual information (right) flow for P3-2 (with $m_{\mathrm{fin}} = 2^{4}$ and $\ell_{\mathrm{fin}} = 2^{6}$).}
\includegraphics[scale=0.29]{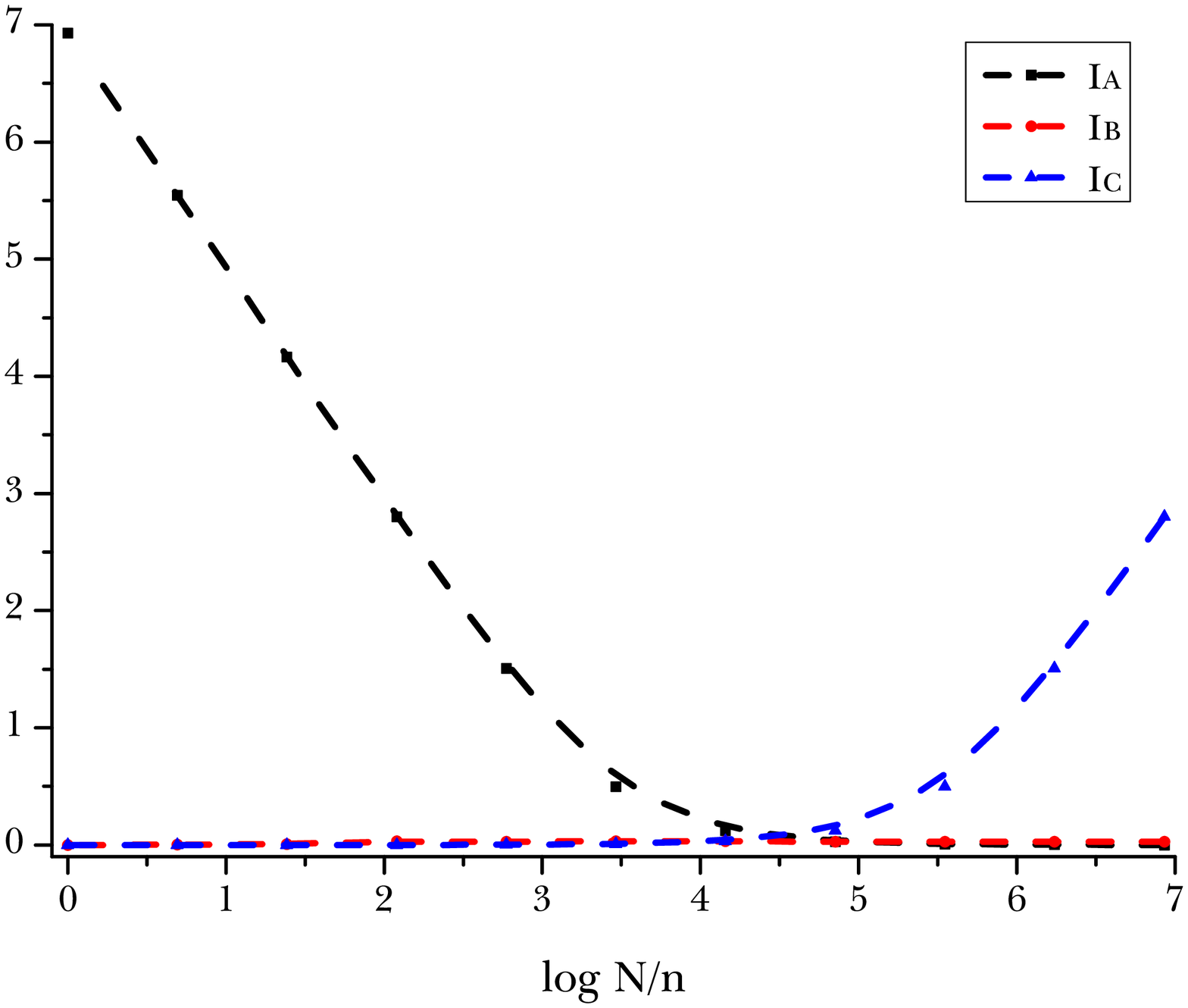}
\includegraphics[scale=0.29]{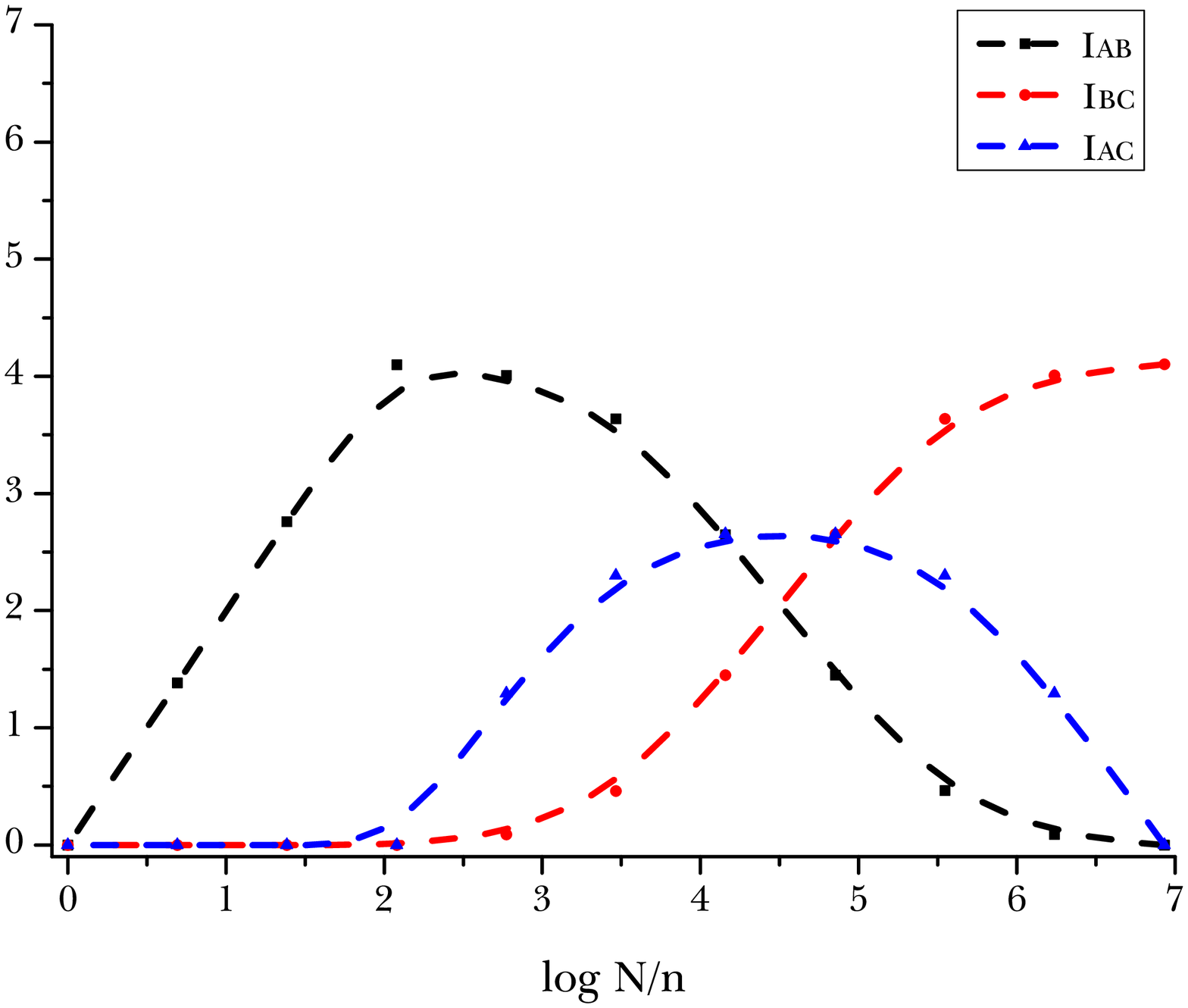}
\caption{\label{fig:P32_37}Information (left) and mutual information (right) flow for P3-2 (with $m_{\mathrm{fin}} = 2^{3}$ and $\ell_{\mathrm{fin}} = 2^{7}$).}
\end{center}
\end{figure}

One remarkable idea to check is the locking of information scenario \cite{Smolin:2005tz}. In this approach, information is stored in the form of the mutual information; in addition, before the very last stage, there is only negligible information in both of inside and outside the black hole. Can this scenario be possible by assuming pure and random states? If the number of states of $C$ (i.e., $\ell_{\mathrm{fin}}$) is smaller than that of $B$ (i.e., $m_{\mathrm{fin}}$), then $I_{B}$ should be greater than zero, and hence the complete realization of the locking of information is not possible. One exception is the case when $m_{\mathrm{fin}} = \ell_{\mathrm{fin}}$ (FIG.~\ref{fig:P31_55}). In this simulation, we take out particles from $A$ to $B$ and $C$ symmetrically and alternatively; then both of $I_{B}$ and $I_{C}$ are almost negligible, while almost all information is stored in the form of the mutual information. In conclusion, the locking of information is realizable only for a very specific case; the number of states of Hawking radiation and that of inside the black hole should be symmetric \textit{until the end of the evaporation}, while it requires a huge black hole interior entropy and hence it seems not very natural in the usual sense. Perhaps, one more possibility is to assume a non-random state, although it needs an independent justification.

\subsubsection{P3-2: Information stored by internal space, bubble universe, bag-of-gold, etc.}

Finally, in the scenario P3-2, we are considering the situation that $m_{\mathrm{fin}} \lesssim \ell_{\mathrm{fin}}$. As we observed in the previous subsection, in the limiting case $m_{\mathrm{fin}} = \ell_{\mathrm{fin}}$ (FIG.~\ref{fig:P31_55}), the behavior of $I_{AB}$ and $I_{AC}$ are almost the same, since $B$ and $C$ are approximately symmetric.

In order to demonstrate the case $m_{\mathrm{fin}} < \ell_{\mathrm{fin}}$, we linearly increased $\ell$ after a certain step, but it is also fair to say that this process mimics a gradual increase of $C$ as well as a continuous leak of information toward a huge space (e.g., a bubble universe \cite{Farhi:1989yr}). In FIGs.~\ref{fig:P32_46} or \ref{fig:P32_37}, we can see clear behaviors that information is transferred in the form of $I_{C}$ and $I_{BC}$. In addition, information of Hawking radiation ($I_{B}$) is almost negligible, and hence Hawking radiation would be almost perfectly thermal.

Therefore, oppositely, if Hawking radiation contains no information (i.e., $I_{B} \simeq 0$) and is completely thermal, then the only way to store information can be found in the scenario P3-2. In this case, the important assumption is $m_{\mathrm{fin}} \lesssim \ell_{\mathrm{fin}}$; so, there should be a hidden and a large number of degrees of freedom. On the other hand, if there is no such a huge space, then Hawking radiation should contain information even with a small amount.

One more interesting observation is that there is a kind of the Page time for $C$: as we can see in FIGs.~\ref{fig:P32_46} or \ref{fig:P32_37}, as the mutual information between $A$ and $C$ reaches its maximum (hence, the entanglements between $A$ and $C$ reach the maximum), $I_{C}$ begins to increase from zero. We can see this behavior since the size of $C$ is greater than that of $B$. This Page time is the time scale that information begins to be transferred to $C$ (e.g., a bubble universe).

\begin{table}
\begin{center}
\begin{tabular}{c| r @{$\gg$} l}
\hline\hline
Cases & \multicolumn{2}{c}{Information contents}\\
\hline
\;\;\;\;\;\;\;\;\;\;$m_{\mathrm{fin}} > \ell_{\mathrm{fin}}$\;\;\;\;\;\;\;\;\;\; & \;\;\;\;\;\;\;\;\;\;$I_{B}, I_{BC}$\;\;\;\; & \;\;\;\;$I_{C}$\;\;\;\;\;\;\;\;\;\; \\
\;\;\;\;\;\;\;\;\;\;$m_{\mathrm{fin}} \simeq \ell_{\mathrm{fin}}$\;\;\;\;\;\;\;\;\;\; & \;\;\;\;\;\;\;\;\;\;$I_{BC}$\;\;\;\; & \;\;\;\;$I_{B}, I_{C}$\;\;\;\;\;\;\;\;\;\; \\
\;\;\;\;\;\;\;\;\;\;$m_{\mathrm{fin}} < \ell_{\mathrm{fin}}$\;\;\;\;\;\;\;\;\;\; & \;\;\;\;\;\;\;\;\;\;$I_{C}, I_{BC}$\;\;\;\; & \;\;\;\;$I_{B}$\;\;\;\;\;\;\;\;\;\; \\
\hline\hline
\end{tabular}
\caption{\label{tab:sum}Summary of the results at the end stage of the evaporation.}
\end{center}
\end{table}

\subsection{Information at the end of the evolution}

Finally, we summarize the most important feature from these experiments. After the black hole emits all particles, the amount of information depends on $m_{\mathrm{fin}}$ (the degrees of freedom of $B$) and $\ell_{\mathrm{fin}}$ (the degrees of freedom of $C$). There are three cases (TABLE~\ref{tab:sum}): (1) $m_{\mathrm{fin}} > \ell_{\mathrm{fin}}$, where information is stored by $I_{B}$ and $I_{BC}$, while $I_{C}$ is negligible; (2) $m_{\mathrm{fin}} \simeq \ell_{\mathrm{fin}}$, where information is stored by $I_{BC}$, while $I_{B}$ and $I_{C}$ are both negligible; and (3) $m_{\mathrm{fin}} < \ell_{\mathrm{fin}}$, where information is stored by $I_{C}$ and $I_{BC}$, while $I_{B}$ is negligible.

This conclusion is not so surprising. At the end of the evaporation, this is nothing but to divide a system to two entangled subsystems. Therefore, as we see $B$ and $C$ at the last stage, we would see the information of the bipartite system such as in FIG.~\ref{fig:bipart}. In this bipartite system, around the marginal limit $m_{\mathrm{fin}} \simeq \ell_{\mathrm{fin}}$ (corresponding to the Page time), the mutual information ($=2S(B|C)$) reaches its maximum, while $I_{B}$ and $I_{C}$ both are negligible. Before this limit (i.e., $m_{\mathrm{fin}} > \ell_{\mathrm{fin}}$, as $m_{\mathrm{fin}}$ decreases, $I_{B}$ decreases and $I_{BC}$ increases, while $I_{C}$ is almost negligible; the opposite thing happens for $m_{\mathrm{fin}} < \ell_{\mathrm{fin}}$. At the end of the time evolution, our results for tripartite systems are exactly consistent with those of bipartite systems.

\section{\label{sec:dis}Discussion}

In this paper, we investigated information flow of tripartite systems. Especially, we focused on three scenarios: (P2) non-local effects continuously transports information, (P3-1) information is stored in the last stage, and (P3-2) information is transferred to the internal huge space.

Although these categories are quite simplified, our numerical investigations give important wisdoms and we could observe various interesting features for each scenario. Especially, we summarize the following important conclusions:
\begin{itemize}
\item[--] If there is a hidden part $C$, then the role of mutual information is very important because a significant amount of information will be stored in the form of $I_{BC}$.
\item[--] When there is a leak of information to a hidden degree of freedom $C$, the entanglement entropy or mutual information sensitively depends on the detailed leakage process. If the leakage happens only the end of the evaporation (P3-1), then during the evaporation we cannot distinguish the information leakage; if it happens gradually (P2), then we can in principle experimentally notice this.
\item[--] If Hawking radiation is completely thermal, then the only consistent way to explain unitarity is to assume that the hidden space $C$ has very large degrees of freedom (on the order of the black hole entropy).
\item[--] There is the marginal limit that realizes the locking of information scenario, but this requires a large entropy of the black hole interior by the end of the evaporation.
\end{itemize}

One significant conclusion from our study is as follows. \textit{The following three contents cannot be consistent}: (1)~unitarity, (2)~Hawking radiation is totally thermal, and (3)~the remnant has small degrees of freedom compared to the initial entropy. This seriously restricts various scenarios of the remnant picture; the existence of the remnant is not enough and we need to explain the huge entropy inside of it.

If it is not possible to imagine (or at least, if it is not generic) such a huge number of degrees of freedom of a remnant, a large volume, or a bubble universe (or, very active non-local transfer of information), then Hawking radiation should contain a bit of information whatever the end stage of the black hole is. In this case, we need to struggle with the menace of black hole complementarity. One may introduce a firewall, but this can cause problems.

Therefore, our work clarified the \textit{sharp tension} between various assumptions regarding the information loss problem. Then do we need to begin to think the possibility that information is not attached by Hawking radiation nor retained by another object, and hence effectively disappeared, but the total information is conserved by a certain way? Perhaps, this kind of the effective loss of information would be a good idea and we remain this for future investigations.

\section*{Acknowledgment}

DY was supported by Leung Center for Cosmology and Particle Astrophysics (LeCosPA) of National Taiwan University (103R4000). This work was supported by the DGIST Undergraduate Group Research Project (UGRP) grant.

\end{document}